\documentclass[epj]{svjour}
\usepackage{graphics}
\usepackage{epsfig}
\usepackage[english]{babel}
\def\sl#1{\slash{\hspace{-0.2 truecm}#1}}
\title{Invariant Amplitudes for Pion Electroproduction}
\author{B. Pasquini\inst{1}, D. Drechsel\inst{2}, L. Tiator\inst{2}
}
\authorrunning{B. Pasquini et al.}%
\institute{Dipartimento di Fisica Nucleare e Teorica, Universit\'a degli Studi
di Pavia; INFN, Sezione di Pavia, Pavia, Italy \and Institut f\"ur Kernphysik,
Johannes Gutenberg-Universit\"at, D-55099 Mainz }
\date{12/Dec/2007}
\begin{document}
\abstract {The invariant amplitudes for pion electroproduction on
the nucleon are evaluated by dispersion relations at constant $t$
with MAID as input for the imaginary parts of these amplitudes. In
the threshold region these amplitudes are confronted with the
predictions of several low-energy theorems derived in the soft-pion
limit. In general agreement with Chiral Perturbation Theory, the
dispersive approach yields large corrections to these theorems
because of the finite pion mass. } \PACS{ {11.55.Fv}, {11.55.Hx},
{13.40.Gp}, {13.60.Le}, {14.20.Gk}, {25.20.Lj}, {25.30.Rw}}
\maketitle
\section{Introduction}

In two recent publications we have studied pion photoproduction on the nucleon
in the framework of fixed-$t$ dispersion relations~\cite{Pas05,Pas06}. In
particular, we have concentrated on the threshold region in which the results
can be compared to both precision data and predictions of baryon chiral
perturbation theory (ChPT). The dispersion relations (DRs) are based on a set
of 4 photoproduction amplitudes $A_i(\nu ,t)$ depending on energy and momentum
transfer described by the Lorentz invariant variables $\nu$  and $t$,
respectively. These relations are Lorentz and gauge invariant by construction,
and unitarity is implemented by constructing the real parts of the amplitudes
from the imaginary (absorptive) parts via the dispersion
integrals~\cite{Den61,Bal61,Ber67,vGe69}. It is possible to evaluate these
amplitudes also outside of the physical region by analytic continuation. In
particular, the dispersive amplitudes for sub-threshold kinematics are regular
functions in a region of small $\nu$ and $t$ values, and therefore they can be
expanded in a power series about the origin of the Mandelstam plane ($\nu =0
,\, t=0$). Comparing this series with the tree and loop contributions of
relativistic baryon ChPT~\cite{Ber92a,Ber94,Ber02,Ber05,Lehn07} one can read off the
required low-energy constants (LECs) of that field theory, which up to now have
been fixed by resonance saturation models or fits to the threshold data. In our
present work we use MAID05~\cite{MAID} as input for the absorptive parts of the
amplitudes, which are obtained over the full resonance region up to
center-of-mass (c.m.) energies of W=2.2~GeV by a global fit to the pion
photoproduction data. With few exceptions the results compare favorably with
the experimental data and the predictions of ChPT in the threshold region.\\

Another interesting aspect is the comparison with sum rules and low-energy
theorems (LETs) of the 1950's and 1960's~\cite{Kro54,Vai70,deB70}, which were
based on current algebra and the PCAC hypothesis (partial conservation of the
axial current). These relations become exact in the chiral limit of QCD, and
thus all variables and observables have to be understood in the fictitious
limit of vanishing (light) quark masses and hence soft pions with mass $M_{\pi}
\rightarrow 0$. In this limit the threshold for pion photoproduction moves to
the origin of the Mandelstam plane ($\nu =0 \,, t=0$). However, extensive
investigations in ChPT~\cite{Ber91,Ber92b,Ber92c,Ber00} have shown that the
finite pion mass leads to substantial corrections at physical threshold ($ \nu
= \nu_{\rm {thr}} \approx 0.136$~GeV, $t = t_{\rm {thr}} \approx
-0.016$~GeV$^2$) and that even a previously derived LET for neutral pion
photoproduction had to be corrected because of the non-analytic structure of
the expansion coefficients in $M_{\pi}$. We have studied this aspect in the
context of the sum rule of Fubini, Furlan, and Rossetti (FFR)~\cite{Fub65},
which relates the nucleon's anomalous magnetic moment $\kappa_N$ to a
dispersion integral over the first pion photoproduction amplitude $A_1 (\nu ,
t)$. Generalized to electroproduction, this integral takes the form
\begin{eqnarray}
F_2^N(Q^2)&& \tau_3 + \Delta_1^N (\nu, t_{\rm {thr}},Q^2)= \label{eq:FFRQ2}\\
&& \frac{4 M_N^2}{\pi e g_{\pi N}} {\cal P}\int_{\nu_{\rm {thr}}}^{\infty}{\rm
d}\nu' \, \frac{\nu'\,{\rm Im}\,A_1^{(N, \pi^0)}(\nu',t_{\rm
{thr}},Q^2)}{\nu'^2-\nu^2}\,,\nonumber
\end{eqnarray}
where $F_2^N(Q^2)$ is the Pauli form factor normalized to
$F_2^N(0)=\kappa_N$. Furthermore, $M_N$ is the mass of the nucleon,
$g_{\pi N}$ the pion-nucleon coupling constant, and $e$ the
elementary charge. Plotted as function of $\nu$ and at $Q^2=0$, the
right-hand side (rhs) of Eq.~(\ref{eq:FFRQ2}) yields a pronounced
Wigner cusp with a maximum of about 2.5 for the proton, about 50\%
higher than the anomalous magnetic moment $\kappa_p$. The origin of
this cusp is the strong $(\pi^+,\pi^0)$ rescattering leading to a
large imaginary part of the S-wave multipole, which opens like a
square root and therefore yields a singularity of the integrand at
the charged pion threshold. If however the integral is evaluated at
$\nu \approx 0$, the loop effect at threshold is no longer enhanced,
and the bulk contribution of the integral stems from the resonance
region, in particular from the $N\rightarrow\Delta(1232)$
transition. Although a decrease of the integral for $\nu \rightarrow
0$ is therefore expected, we were surprised that near the origin of
the Mandelstam plane the "FFR discrepancy" $\Delta_1^N (\nu, t, 0)$
is actually compatible with zero, in agreement with the FFR sum
rule. However, this sum rule is derived for a world of massless
pions, which would not only lower the threshold to zero but also
change the anomalous magnetic moment and
the absorptive spectrum.\\

In the present contribution we extend our work to the electroproduction of
pions, which involves two additional longitudinal amplitudes and one additional
variable, the virtuality $Q^2$ of the exchanged photon. Moreover, the threshold
now depends on $Q^2$, i.e., $\nu_{\rm {thr}}=\nu_{\rm {thr}}(Q^2)$ and $t_{\rm
{thr}}=t_{\rm {thr}}(Q^2)$. As in the real photon case, the threshold region
opens a wide field of comparisons with recent experiments and predictions of
baryon ChPT. For example, the loop corrections of ChPT have a very distinct
$Q^2$ dependence~\cite{Ber94}, completely different from the form factors in
the pole contributions, and some of the new experimental data still offer
problems for the theoretical description. As a first step towards a new
dispersive approach, we address two sum rules for virtual photons. The first
one is given by Eq.~(\ref{eq:FFRQ2}) as function of $Q^2$. The second sum rule
connects the axial ($G_A^V$) and Dirac ($F_1^V$) isovector form factors to the
longitudinal amplitude $A_6$ with isospin (-). Its physics content is identical
with the LET of Nambu {\it et al.}~\cite{Nam62}, which has been derived for the
slope of the S-wave multipole. In the notation of Fubini {\emph et
al.}~\cite{Fub65} this sum rule may be cast into the form
\begin{eqnarray}
G_A^V(Q^2)&-& F_1^V(Q^2)+ \Delta_6^{(-)}(\nu, t_{\rm {thr}},Q^2)=
\label{eq:A6FFRQ2} \\
&& \frac {4M_NQ^2}{\pi eg_{\pi N}}\,{\cal P}\int_{\nu_{\rm {thr}}}^{\infty}{\rm
d}\nu'\, \frac{\nu'\,{\rm Im}\,A_6^{(-)}(\nu',t_{\rm thr},Q^2)}
{\nu'^2-\nu^2}\,.\nonumber
\end{eqnarray}
As in the former case, this sum rule is derived in the soft-pion limit, for
which both the kinematic variables and the observables (form factors,
multipoles) differ from the physical ones by terms ${\cal O}(M_{\pi})$. In the
limit $Q^2 \rightarrow 0$, the lhs of Eq.~(\ref{eq:A6FFRQ2}) yields information
on $\langle r^2\rangle_1^V - \langle r^2\rangle_A^V$, where $\langle
r^2\rangle_i^V$ are the squares of the respective root-mean-square (r.m.s.)
radii. The integral on the rhs now involves longitudinal multipoles also at
$Q^2=0$, which requires an extrapolation from the measured values at finite
$Q^2$. Moreover, the two radii are of similar size, and the first
estimates~\cite{Ria66} simply led to the result $G_A^V(Q^2)-F_1^V(Q^2)=0$. The
first and to our knowledge only dispersive calculation of
Eq.~(\ref{eq:A6FFRQ2}) was performed by Adler and Gilman already in
1966~\cite{Adl66}. The result was $\langle r^2\rangle_1^V - \langle
r^2\rangle_A^V=0.152$~fm$^2$, in fantastic agreement with our present knowledge
of this observable, $(0.14 \pm 0.03)$~fm$^2\,$! Unfortunately, it has to be
realized that the multipoles used in performing the integral had large error
bars. In particular, the longitudinal and transverse multipoles were assumed to
be equal, which is only correct for the unphysical kinematics of the Siegert
limit. As an example, the longitudinal and transverse S-wave multipoles take
the same value in that limit, but already at $Q^2=0$ they are quite different,
and in the resonance region even the relative sign between the longitudinal and
the respective transverse multipoles may differ from the low-energy limit.
However, the real merit of this early work is the observation that formidable
cancelations occur (I) among contributions of positive sign in the region up to
the $\Delta (1232)$ resonance and of negative sign in the second resonance
region and (II) between the electric transverse
and longitudinal contributions of the same multipolarity.\\

We proceed by summarizing the kinematics for pion electroproduction in Sec.~2,
and in Sec.~3 we introduce the invariant and CGLN amplitudes. The status of the
LETs and sum rules at finite $Q^2$ is discussed in Sec.~4. We present the
predictions of dispersion theory in Sec.~5 and close by a short summary in
Sec.~6.
\section{Kinematics}
\begin{figure}[htb]
\begin{center}
\epsfig{figure=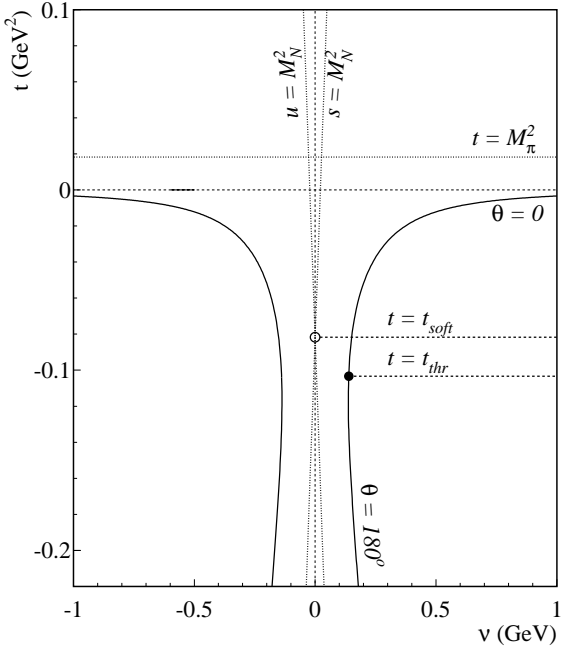,width=8cm,angle=0}
\end{center}
\caption{The Mandelstam plane for pion electroproduction on the nucleon. The
solid line shows the boundary of the physical region for $Q^2=0.1~$GeV$^2$.
This boundary corresponds to forward production ($\theta=0^{\circ}$) for $t \ge
t_{\rm thr}$ and to backward production ($\theta=180^{\circ}$) for $t \le
t_{\rm thr}$. The nucleon and pion pole positions are indicated by the dotted
lines $s=M_N^2$, $u=M_N^2$, and $t=M_{\pi}^2$. The threshold of pion
electroproduction is represented by the solid circle, in the soft-pion limit
the threshold moves to $\nu=\nu_B=0$ (open circle).} \label{fig:mandelstam}
\end{figure}
Let $p_i^\mu $ and $p_f^\mu$ be the four-momenta of the initial and final
nucleons, and $k^\mu$ and $q^\mu$ the four-momenta of the photon and pion,
respectively. In the c.m. system, we define
\begin{eqnarray}
p_i^\mu = (E_i, -{\bf k}), & &p_f^\mu = (E_f, -{\bf q}),\nonumber\\
k^\mu=(k_0, {\bf k}),\quad & & q^\mu=(q_0, {\bf q}) \, .
\end{eqnarray}

The familiar Mandelstam variables are
\begin{eqnarray}
s=(p_i+k)^2,\quad t=(q-k)^2,\quad u=(p_i-q)^2,
\end{eqnarray}
and $\nu=(s-u)/(4M_N)$ is the crossing symmetric variable. This variable is
related to the photon lab energy $E_\gamma^{\rm lab}$ by $\nu=E_\gamma^{\rm
lab} + (t-M_\pi^2+Q^2)/(4M_N)$. The physical $s$-channel region is shown in
Fig.~\ref{fig:mandelstam} for $Q^2=0.1$~GeV$^2$. Its upper and lower boundaries
are given by the scattering angles $\theta=0$ and $\theta=180^{\circ}$,
respectively. The nucleon and pion poles lie in the unphysical region and are
indicated by the dotted lines at $\nu_s=\nu_B$ (s-channel) and $\nu_u=-\nu_B$
(u-channel), where
\begin{eqnarray}
\nu_B =\frac{t-M_{\pi}^2+Q^2}{(4M_N)}\,.
\end{eqnarray}
The threshold for pion electroproduction lies at
\begin{eqnarray}
\nu_{\rm {thr}}
&=& \frac{M_{\pi} [(2M_N+M_{\pi})^2+Q^2]}{4M_N (M_N+M_{\pi})}\,,\nonumber\\
t_{\rm {thr}} &=& -\frac{ M_N(M_{\pi}^2+Q^2)}{M_N+M_{\pi}}\,,\label{nuthr_tthr}
\end{eqnarray}
and the energies and momenta of the particles are given by
\begin{eqnarray}
k_0 &=& \frac{s-Q^2-M_N^2}{2\sqrt{s}},\quad
q_0=\frac{s+M_{\pi}^2-M_N^2}{2\sqrt{s}},\nonumber\\
q &=&|{\bf q}|=\left[\left(\frac{s+M_{\pi}^2-M_N^2}{2\sqrt{s}}\right)^2
-M_{\pi}^2\right]^{1/2},\nonumber\\
k &=&|{\bf k}|=\left[\left(\frac{s-Q^2-M_N^2}{2\sqrt{s}}\right)^2
+Q^2\right]^{1/2},\label{eq:kin}\\
E_i &=& W-k_0=\frac{s+M_N^2+Q^2}{2\sqrt{s}}\,,\nonumber\\
E_f &=& W-q_0=\frac{s+M_N^2-M_{\pi}^2}{2\sqrt{s}}\,,\nonumber
\end{eqnarray}
with $W=\sqrt{s}$ the c.m. energy and $Q^2=-k^\mu k_\mu$. The pseudothreshold
or Siegert limit for electroproduction lies at the unphysical point ${\bf
k}=0$, which corresponds to $Q^2=-(W-M_N)^2$.
\section{Invariant and CGLN Amplitudes}
The electromagnetic transition can be described by 6 invariant amplitudes
$A_i$,
\begin{eqnarray}
\varepsilon_{\mu} \, J^{\mu} = \sum_{i=1}^6 A_i(\nu,t,Q^2)\,\varepsilon_{\mu}\,
M^\mu_i, \label{eq:current}
\end{eqnarray}
with $\varepsilon_{\mu}$ the polarization four-vector of the virtual photon and
$J_{\mu}$ the transition current of the nucleon. In the notation of
Dennery~\cite{Den61}, the four-vectors $M^\mu_i$ take the following form :
\begin{eqnarray}
M^\mu_1&=&
-\frac{1}{2}i\gamma_5\left(\gamma^\mu\sl{k}-\sl{k}\gamma^\mu\right)\, ,
\nonumber\\
M^\mu_2&=&2i\gamma_5\left(P^\mu\, k\cdot(q-\frac{1}{2}k)-
(q-\frac{1}{2}k)^\mu\,k\cdot P\right)\, ,\nonumber\\
M^\mu_3&=&-i\gamma_5\left(\gamma^\mu\, k\cdot q
-\sl{k}q^\mu\right)\, ,\label{eq:M1-6} \\
M^\mu_4&=&-2i\gamma_5\left(\gamma^\mu\, k\cdot P
-\sl{k}P^\mu\right)-2M_N \, M^\mu_1\, ,\nonumber\\
M^\mu_5&=&i\gamma_5\left(k^\mu\, k\cdot q
+Q^2 q^\mu\right)\, ,\nonumber\\
M^\mu_6&=&-i\gamma_5\left(\sl{k}k^\mu+Q^2 \gamma^\mu\right)\, ,\nonumber
\end{eqnarray}
with $P^\mu=(p_i+p_f)^\mu \, /2$, $\sl{a}=a_\mu\gamma^\mu$, and $\gamma$
matrices as defined in Ref.~\cite{Bjo65}. In the case of real photons ($Q^2=0$)
and with the gauge condition $\varepsilon_\mu k^\mu=0$, the matrices $M^\mu_5$
and $M^\mu_6$ do not contribute to the interaction Lagrangian, and the
remaining four matrices reduce to Eq.~(10) of Ref.~\cite{Pas05}. The invariant
amplitudes $A_i$ can be further decomposed into three isospin channels
($a=1,2,3$),
\begin{eqnarray}
A_i^a=A_i^{(-)}i\epsilon^{a3b}\tau^b+A_i^{(0)}\tau^a+A_i^{(+)}\delta_{a3},
\end{eqnarray}
where $\tau^a$ are the Pauli matrices in isospace, and the physical
photoproduction amplitudes are given by
\begin{eqnarray}
A_i(\gamma p\rightarrow n\pi^+)&=&\sqrt{2}(A_i^{(-)}+A_i^{(0)}),\nonumber\\
A_i(\gamma p\rightarrow p\pi^0)&=&A_i^{(+)}+A_i^{(0)},\\
A_i(\gamma n\rightarrow p\pi^-)&=&-\sqrt{2}(A_i^{(-)}-A_i^{(0)}),\nonumber\\
A_i(\gamma n\rightarrow n\pi^0)&=&A_i^{(+)}-A_i^{(0)}.\nonumber
\label{eq:phys_chann}
\end{eqnarray}
Under crossing, the amplitudes $A_{1,2,4}^{(+,0)}$ and $A_{3,5,6}^{(-)}$ are
even functions of $\nu$ and satisfy a DR of the type
\begin{eqnarray}
&&A_{i,{\rm disp}}^{(I)}(\nu,t,Q^2) \equiv {\rm Re}A^{(I)}_i(\nu,t,Q^2) -
A_{i,{\rm pole}}^{(I)}(\nu,t,Q^2)\nonumber\\
&& \quad \quad = \frac{2}{\pi}{\cal P}\int_{\nu_{thr}}^{\infty}{\rm d}\nu'
\frac{\nu'\,{\rm Im}A_i^{(I)}(\nu',t,Q^2)}{\nu'^2-\nu^2}\,,\label{eq:dr1}
\end{eqnarray}
whereas the amplitudes $A_{3,5,6}^{(+,0)}$ and $A_{1,2,4}^{(-)}$ are odd and
therefore fulfil the relation
\begin{equation}
A_{i,{\rm disp}}^{(I)}(\nu,t,Q^2)=\frac{2\nu}{\pi}{\cal P}\int_{\nu_{\rm
thr}}^{\infty}{\rm d}\nu' \frac{{\rm
Im}A_i^{(I)}(\nu',t,Q^2)}{\nu'^2-\nu^2}\,.\label{eq:dr2}
\end{equation}
The nucleon pole contributions $A_{i,{\rm pole}}^{(I)}$ can be written as
functions of the Mandelstam variables and $Q^2$ as follows:
\begin{eqnarray}
A_{1,\,{\rm pole}}^{(I)} & = & \ \ \ \frac{eg_{\pi N}}{2}
\left(\frac{1}{s-M_N^2}+\frac{\epsilon^I}{u-M_N^2}\right)\,F_1^{(I)}(Q^2)
\,,\nonumber \\
A_{2,\,{\rm pole}}^{(I)}  & = & -\frac{eg_{\pi N}}{t-M^2_\pi}
\left(\frac{1}{s-M_N^2}+\frac{\epsilon^I}{u-M_N^2}\right)\,F_1^{(I)}(Q^2)
\,,\nonumber \\
A_{3,\,{\rm pole}}^{(I)} & = & -\frac{eg_{\pi N}}{4M_N}
\left(\frac{1}{s-M_N^2}-\frac{\epsilon^I}{u-M_N^2}\right)\,F_2^{(I)}(Q^2)
\,,\label{eq:a1-6pole}\\
A_{4,\,{\rm pole}}^{(I)}& = & -\frac{eg_{\pi N}}{4M_N}
\left(\frac{1}{s-M_N^2}+\frac{\epsilon^I}{u-M_N^2}\right)\,F_2^{(I)}(Q^2)
\,,\nonumber\\
A_{5,\,{\rm pole}}^{(I)} & = & -\frac{eg_{\pi N}}{2(t-M_{\pi}^2)}\,
\left(\frac{1}{s-M_N^2}-\frac{\epsilon^I}{u-M_N^2}\right)\,F_1^{(I)}(Q^2)
\nonumber \\
&& + \frac{2eg_{\pi N}[F_{\pi} ^V(Q^2)-F_1^{(-)}(Q^2)]}
{Q^2(t-M_{\pi}^2)}\,\delta_{I,-}
\,,\nonumber\\
A_{6,\,{\rm pole}}^{(I)} & = & 0\,, \nonumber
\end{eqnarray}
with $\epsilon^+=\epsilon^0=-\epsilon^-=1$, $F_i^{(0)}=F_i^p+F_i^n=F_i^S$  the
isoscalar and $F_i^{(+,-)}=F_i^p-F_i^n=F_i^V$ the isovector form factors,
normalized to $F_1^I(0)=1$, $F_2^{(0)}=\kappa_p+\kappa_n$, and
$F_2^{(+,-)}=\kappa_p-\kappa_n$, where $\kappa_p$ and $\kappa_n$ are the
anomalous magnetic moments of proton and neutron, respectively.\\

For further use we also list the contributions of $t$-channel vector-meson
exchange to the isospin (+) and (0) amplitudes :
\begin{eqnarray}
A_1^{(+,0)}(t,Q^2) & = & \frac{e\lambda_V\,g_V^{(T)}}{2M_NM_\pi}
\frac{t}{t-m_V^2} \, F_{\gamma \pi V}(Q^2)\,,
\nonumber \\
A_2^{(+,0)}(t,Q^2) & = & -\frac{e\lambda_V\,g_V^{(T)}}{2M_NM_\pi} \,
\frac{t-M_{\pi}^2-Q^2}{(t-m_V^2) \, (t-M_{\pi}^2)} \, F_{\gamma \pi V}(Q^2)\,,
\nonumber \\
A_3^{(+,0)}(t,Q^2)& = & 0 \,, \label{eq:a1-6VM}\\
A_4^{(+,0)}(t,Q^2) & = & - \frac{e\lambda_V\,g_V^{(V)}}{M_\pi}
\frac{1}{t-m_V^2} \, F_{\gamma \pi V}(Q^2)\,,
\nonumber \\
A_5^{(+,0)}(t,Q^2) & = & \frac{e\lambda_V\,g_V^{(T)}}{M_\pi} \,
\frac{\nu}{(t-m_V^2) \, (t-M_{\pi}^2)}\, F_{\gamma \pi V}(Q^2)\, ,
\nonumber \\
A_6^{(+,0)}(t,Q^2) & = & 0 \,,\nonumber
\end{eqnarray}
where $\lambda_V$ denotes the coupling of the vector meson ($V=\omega,\rho$) to
the $\gamma\pi$ system, $g_V^{(V,T)}$ its vector or tensor coupling to the
nucleon, and $F_{\gamma \pi V}(Q^2)$ is a transition form factor. For further
details see Ref.~\cite{Pas06}.\\

The matrix element of the electromagnetic current, Eq.~(\ref{eq:current}),
takes the form
\begin{eqnarray}
\bar u(p_f) \, \sum_{i=1}^{6} \,A_i\, \varepsilon_\mu M_i^\mu\, u(p_i)
=\frac{4\pi W}{M_N}\chi_f^\dagger {\cal F} \chi_i\,, \label{eq:matrix}
\end{eqnarray}
with $u(p)$ the Dirac spinor of the nucleon with $\bar u(p)u(p)=2 M_N$, and
$\chi$ the Pauli spinor of the nucleon. The operator ${\cal F}$ in
Eq.~(\ref{eq:matrix}) can be decomposed into the CGLN amplitudes ${\cal
F}_i$~\cite{Che57},
\begin{eqnarray}
{\cal F} &=& -i\,({\vec {\sigma}}\cdot{\bf {b}}) \, {\cal F}_1 -\,({\vec
{\sigma}} \cdot\hat {\bf {q}})\, {\bf {b}}\cdot({\vec {\sigma}} \times \hat{\bf
{k}})\,
{\cal F}_2 -\nonumber\\
&i&\,({\bf {b}}\cdot\hat {\bf {q}})\, ({\vec {\sigma}} \cdot\hat {\bf {k}})
{\cal F}_3 - i ({\bf {b}} \cdot \hat{\bf {q}})({\vec {\sigma}} \cdot \hat {\bf
{q}}) \, {\cal F}_4 + \label{eq:emcurrent}\\
&i&\,({\vec {\sigma}}\cdot\hat {\bf{k}})\, b_0\frac{k}{k_0} {\cal F}_5 +
i\,({\vec {\sigma}}\cdot\hat {\bf {q}})\, b_0\frac{k}{k_0} {\cal F}_6 \,
,\nonumber
\end{eqnarray}
where $b^\mu=\varepsilon^\mu - ({\vec {\epsilon}}\cdot \hat {\bf {k}})
k^\mu/k$. The relations between the invariant amplitudes $A_i$ and the CGLN
amplitudes ${\cal F}_i$ are obtained by combining Eqs.~(\ref{eq:matrix}) and
(\ref{eq:emcurrent}). The general result is given in Appendix~\ref{app:AtoF}
and the multipole series for the CGLN amplitudes is shown in
Appendix~\ref{app:FtoM}. Specifically, the invariant amplitude $A_1$ at
threshold has the following multipole decomposition:
\begin{eqnarray}
A_1^{\rm thr} & = &\frac{4\pi}{M_N\,(\mu^2+\rho)\sqrt{(1+\mu)\,[(2+\mu)^2+\rho]}}\nonumber\\
& \times & \bigg\{(1 + \mu)[\mu(2+\mu)+\rho] \, E_{0+}+\frac{4(1+\mu)^2 \,
\rho}{\mu(2+\mu)-\rho}\,L_{0+} \nonumber\\
& - & M_N\,\mu\,\sqrt{(\mu^2+\rho)\,[(2+\mu)^2+\rho]}\,\bar{P}_2\label{eq:thresh_A1} \\
& - & \frac{M_N (2+\mu)\,(\mu^2+\rho)^{3/2}} {\sqrt{(2+\mu)^2+\rho}}
\,\bar{P}_3 - 2M_N^2\,\mu\,(\mu^2+\rho)\,\bar{D}\nonumber \\
& + & \frac{8 M_N \,(1+\mu)^2 \,\rho \sqrt{\mu^2+\rho}}{[\mu(2+\mu)-\rho]\,
\sqrt{(2+\mu)^2+\rho}}\,\bar{P}_5\bigg\}\,,\nonumber
\end{eqnarray}
where we have introduced the ratios $\mu=M_\pi/M_N$ and $\rho=Q^2/M_N^2$, and
the following combinations of the P- and D-wave multipoles:
\begin{eqnarray}
\bar{P}_2 & = & (3E_{1+}-M_{1+}+M_{1-})/q \,,\nonumber\\
\bar{P}_3 & = & (2M_{1+}+M_{1-})/q \,, \label{eq:wave_combi} \\
\bar{P}_5 & = & (L_{1-}-2L_{1+})/q \,,\nonumber\\
\bar{D} & = & 3(M_{2+}-E_{2+}-M_{2-}-E_{2-})/q^2 \,.\nonumber
\end{eqnarray}
We note that all the multipoles in Eqs.~(\ref{eq:thresh_A1}) to
(\ref{eq:thresh_A1_alt}) should be evaluated at $W_{\rm thr}=M_N (1+ \mu)$. The
factor $(\mu^2+\rho)^{-1}$ in Eq.~(\ref{eq:thresh_A1}) is worrying because of
the singularity in the Siegert limit $k \rightarrow 0$, which corresponds to
$Q^2 \rightarrow -M_{\pi}^2$ or $\rho \rightarrow - \mu^2$ in our notation. We
may eliminate this critical factor by use of the Siegert limit for the
multipoles (see Appendix~\ref{app:FtoM}). In particular, the S-wave multipoles
are even functions of $k$, and since both $L_{0+}$ and $E_{0+}$ approach the
same constant in that limit, the difference of the two multipoles is
proportional to $k^2$. Furthermore, all the P-waves contain a factor $k$. With
the definitions $\Delta_{0+}=(L_{0+} - E_{0+})/k^2$ and ${\cal
P}_i=P_i/(q\,k)=\bar{P}_i/k$ we can rewrite Eq.~(\ref{eq:thresh_A1}) as
follows:
\begin{eqnarray}
A_1^{\rm thr} & = & \frac{4\pi M_N}{\sqrt{(1+\mu)\,[(2+\mu)^2+\rho]}}\nonumber\\
& \times & \bigg\{ \frac{(1 + \mu)[(2+\mu)^2-\rho]}{M_N^2 \, (\mu(2+\mu)-\rho)}
\, E_{0+}+\frac{[(2+\mu)^2+\rho] \, \rho}{\mu(2+\mu)-\rho}\,
\Delta_{0+} \nonumber\\
& - & \frac{ \mu \,[(2+\mu)^2+\rho]}{2(1+\mu)}\,{\cal P}_2-
\frac{(2+\mu)\,(\mu^2+\rho)}{2(1+\mu)}\,{\cal P}_3\label{eq:thresh_A1_alt} \\
& - &  2\mu\,\bar{D} + \, \frac{4 \,(1+\mu) \,\rho}{\mu(2+\mu)-\rho}\,{\cal
P}_5 \bigg\} \,.\nonumber
\end{eqnarray}

Note that we have left the term $\bar{D}$ unchanged, because only the first
three D waves in Eq.~(\ref{eq:wave_combi}) are proportional to $k^2$ in the
Siegert limit, whereas $E_{2-}$ is an electric dipole transition and therefore
approaches a constant in that limit. Comparing now Eq.~(\ref{eq:thresh_A1})
with Eq.~(\ref{eq:thresh_A1_alt}), we find that the pole at $(\mu^2+\rho)$ has
disappeared in the latter equation. The remaining kinematical pole at
$\mu(2+\mu)-\rho$ is compensated by a zero in the longitudinal multipoles. The
corresponding equations for the longitudinal amplitudes $A_5$ and $A_6$ are
given in Appendix~\ref{app:A5A6}.\\

It is evident from Eq.~(\ref{eq:M1-6}) that the 6 ``Dennery amplitudes'' $A_i$
fulfill gauge invariance. However, Eqs.~(\ref{eq:inv_A2b}) and
(\ref{eq:inv_A5b}) of Appendix~\ref{app:AtoF} show that the amplitudes $A_2$
and $A_5$ have kinematical singularities at $t=M_\pi^2$. These singularities
can be avoided by introducing a set of 8 four-vectors $N_i^\mu$~\cite{Bal61},
which are free of kinematical singularities and therefore should obey a
Mandelstam representation. However, these amplitudes are not separately gauge
invariant. In order to implement gauge invariance, the associated ``Ball
amplitudes'' $B_i$ have to fulfill two additional constraints~\cite{Bal61}. As
discussed in more detail by v.~Gehlen~\cite{vGe69}, these constraints lead to
an additional kinematical singularity at the pion pole ($t=M_\pi^2$), which
should not contribute to the residue of (I) $A_2$ for all values of $\nu$ and
real photons ($Q^2=0$), and (II) $A_5$ for $\nu=0$ and all values of $Q^2$.
This has no further consequences for the function $A_2$. However, the insertion
of the second condition in Eq.~(\ref{eq:dr1}) requires that
\begin{equation}
\label{eq:A_5} {\rm Re}\,A^{(-)}_{5,\, {\rm disp}}(0,t,Q^2)= \frac{2}{\pi}
\int_{\nu_{thr}}^\infty \frac{{\rm d}\nu'}
{\nu'}\,{\rm{Im}}\,A_5^{(-)}(\nu',t,Q^2)\,.
\end{equation}
Since the invariant amplitudes should only have a dynamical singularity in $t$,
namely the pion pole term, any contribution of the dispersive integral with the
behavior of the pion pole term has to be subtracted, i.e., Eq.~(\ref{eq:dr1})
has to be corrected by
\begin{eqnarray}
&&{\rm Re}\,A^{(-)}_5(\nu,t,Q^2) \rightarrow {\rm Re}A^{(-)}_5(\nu,t,Q^2)- \label{eq:A_5corr} \\
&& \frac{2}{\pi}\ \frac{1}{t-M_{\pi}^2}\int_{\nu_{\rm thr}}^{\infty} \frac{{\rm
d}\nu'}{\nu'}\,\lim_{t'\rightarrow
M_{\pi}^2}\,\{(t'-M_{\pi}^2)\,{\rm{Im}}A_5^{(-)}(\nu',t',Q^2)\}\,.\nonumber
\end{eqnarray}
As a result the corrected dispersion integral does no longer contribute to the
pion-pole residue. In particular at $\nu=0$, this residue is given by the pole
contribution of Eq.~(\ref{eq:a1-6pole}) as follows:
\begin{equation}
\label{eq:A_5_pole} A_{5, {\rm pole}}^{(-)}(0,t,Q^2) = \frac{2eg_{\pi
N}}{Q^2}\,\left\{ \frac{F_\pi^V(Q^2)}{t-M_\pi^2} - \frac{F_1^V
(Q^2)}{t-M_\pi^2+Q^2} \right\}\,.
\end{equation}
\section{Low Energy Theorems and Sum Rules}
Several LETs for pion photo- and electroproduction were derived in the 1950's
and 1960's from PCAC and current algebra which preceded QCD. A modern framework
to derive these sum rules is provided by ChPT as an effective realization of
QCD in terms of its low-energy degrees of freedom. The first of these theorems
is due to Kroll and Ruderman~\cite{Kro54} who found that the threshold
photoproduction of charged pions is described by minimal coupling of the photon
to the pseudovector pion-nucleon interaction. Nambu and
collaborators~\cite{Nam62} extended these considerations to virtual photons and
obtained a relation between the S-wave multipole for pion electroproduction
with isospin $(-)$ and the difference of the isovector Dirac and axial form
factors. Finally Fubini, Furlan, and Rossetti~\cite{Fub65} derived a sum rule
for the Pauli form factors in terms of the amplitudes $A_1^{(+,0)}$. The origin
of these sum rules is summarized as follows:
\begin{enumerate}
\item[(I)]
The electromagnetic transition current $J_\mu$ can be expanded in the set of
the 6 covariants $M_i^\mu$, see Eq.~(\ref{eq:M1-6}). In the limit of vanishing
pion four-momentum only the covariants $M_1^\mu$ and $M_6^\mu$ survive, and
therefore the soft-pion transition current is completely determined by the
associated invariant amplitudes $A_1$ and $A_6$.
\item[(II)]
The soft-pion limit $q_\mu\to0$ is obtained by first going to threshold (${\bf
q}=0, \, q_0=M_\pi$ in the hadronic c.m. system) and then removing the mass of
the pion ($q_0=M_\pi\to 0$). The second step requires an extrapolation into
unphysical territory, which can only be performed within a theoretical
framework.
\end{enumerate}
In the following we consider the dispersive contribution to threshold
amplitudes, $A_{i,\,{\rm disp}}^{(I){\rm thr}}\equiv A_{i,\,{\rm
disp}}^{(I)}(\nu_{\rm thr},t_{\rm thr},Q^2)$, where $\nu_{\rm thr}=\nu_{\rm
thr}(Q^2)$ and $t_{\rm thr}=t_{\rm thr}(Q^2)$ are given by
Eq.~(\ref{nuthr_tthr}). According to Ref.~\cite{Fub65}, the crossing-even
amplitudes $A_{1,\,{\rm disp}}^{(+,0),\,{\rm thr}}$ and $A_{6,\,{\rm
disp}}^{(-),\,{\rm thr}}$ take the following form in the soft-pion limit:
\begin{eqnarray}
A_{1,\,{\rm disp}}^{(+,0){\rm thr}} & \longrightarrow &
\frac{eg_{\pi N}}{4M_N^2}\,F_2^{V,S}(Q^2)\equiv A_{1,{\rm FFR}}^{(+,0)}(Q^2)\,, \label{eq:A1FFR} \\
A_{6,\,{\rm disp}}^{(-){\rm thr}} & \longrightarrow & \frac{eg_{\pi
N}}{2M_NQ^2}\,[G_A^V(Q^2)-F_1^V(Q^2)]\equiv
A_{6,{\rm FFR}}^{(-)}(Q^2)\nonumber\\
& = &  \frac{eg_{\pi N}}{12M_N}\,\left( \, \langle r^2\rangle_1^V-\langle
r^2\rangle_A^V \, \right)+ \ldots\,,\label{eq:A6FFR}
\end{eqnarray}
whereas the crossing-odd amplitudes $A_1^{(-)}$ and $A_6^{(+,0)}$ vanish in
that limit. We repeat that these results are only valid in the world of
massless pions.\\

The corrections due to the finite mass of the pion have been calculated in
ChPT. For further use we list the S-wave multipoles at threshold as obtained to
${\cal O}(q^2)$ by manifestly Lorentz invariant baryon ChPT~\cite{Ber94},
\begin{eqnarray}
E_{0+}^{(+){\rm thr}} & = & x_{\pi N} \bigg\{-2 \mu +
\left(3+\kappa_V\right)\mu^2 + (1 + \kappa_V)\rho \nonumber \\
&+& 2 \mu^2 y_{\pi N} \Xi_1\bigg\}, \nonumber \\
L_{0+}^{(+){\rm thr}}  & = & E_{0+}^{(+){\rm thr}}+  x_{\pi N}
(\mu^2+\rho)\bigg\{- \kappa_V + 2 y_{\pi N} \Xi_2 \bigg
\},\nonumber\\
E_{0+}^{(0){\rm thr}}  & = &  x_{\pi N}  \bigg\{-2 \mu +
\left(3+\kappa_S\right)\mu^2 +(1 + \kappa_S) \rho \bigg\}\,,\nonumber\\
L_{0+}^{(0){\rm thr}}  & = & E_{0+}^{(0){\rm thr}}- x_{\pi N}
(\mu^2+\rho)\,\kappa_S , \label{eq:ChPT}  \\
E_{0+}^{(-){\rm thr}} & = & 4  x_{\pi N} \bigg\{ 1-\mu +
({\textstyle{\frac{9}{8}}}+C_0)\mu^2 \nonumber \\
& -&
\frac{\rho}{4}\left[\kappa_V+{\textstyle{\frac{1}{2}}}+{\textstyle{\frac{2}{3}}}M_N^2
\langle r^2\rangle_A^V
\right] +  \frac{\mu^2 y_{\pi N}}{\pi} \Xi_3\bigg\} ,\nonumber \\
L_{0+}^{(-){\rm thr}}  & = & E_{0+}^{(-){\rm thr}} + x_{\pi
N}(\mu^2+\rho)\bigg\{\kappa_V \nonumber \\
&-& \frac{2\sqrt{(2+\mu)^2+\rho}} {(1+\mu)^{3/2}[\mu^2(2+\mu) + \rho]} +
{\textstyle{\frac{2}{3}}}M_N^2\langle r^2\rangle_A^V \nonumber \\
&-&{\textstyle{\frac{2}{3}}}M_N^2\left[1- \frac{\rho}{2\mu^2+\rho}
\right]\langle r^2\rangle_{\pi}^V + \frac{4\,y_{\pi N}}{\pi}\Xi_4 \bigg \}
,\nonumber
\end{eqnarray}
with $C_0\approx -0.725$ as estimated by resonance saturation, $x_{\pi
N}=eg_{\pi N}/(32\pi M_N)$, and $y_{\pi N}=M_N^2/(8\pi F_{\pi}^2)$.
Furthermore, the loop functions $\Xi_i=\Xi_i( Q^2/M_{\pi}^2)$ can be expanded
in a power series for $Q^2 \ll M_{\pi}^2$,
\begin{eqnarray}
\Xi_1(Q^2/M_{\pi}^2)
&=& \pi+\frac{Q^2}{2M_\pi^2} (4- \pi) + \ldots\,, \nonumber\\
\Xi_2(Q^2/M_{\pi}^2)
&=& (2-\pi)+ \frac{2 Q^2}{M_\pi^2}(\pi - 3) + \ldots\,, \label{eq:Xi}\\
\Xi_3(Q^2/M_{\pi}^2)
&=&\frac{\pi^2}{8}+\frac{1}{2}+\frac{Q^2}{16M_\pi^2}\,(12-\pi^2) + \ldots\,,\nonumber\\
\Xi_4(Q^2/M_{\pi}^2) &=& \frac{1}{8}(4-\pi^2) + \frac{Q^2}{4M_\pi^2} (\pi^2-9)
+ \ldots\,.\nonumber
\end{eqnarray}
This expansion shows that the loop contributions increase with $Q^2$, whereas
the form factors in the pole and FFR terms generally decrease with the
virtuality of the photon. Although the loop effects of Eq.~(\ref{eq:ChPT}) are
formally of ${\cal O} (q^2)$, they yield large corrections to the slopes in
$Q^2$. In the case of neutral pion photoproduction, with isospin (+) and (0),
these effects are even more important such that the LET of the 1960's had to be
revised~\cite{Ber91,Ber92a}. In fact, the lowest order loop corrections for
neutral pion photoproduction have a larger absolute value than the leading
term.\\

Equation~(\ref{eq:ChPT}) contains the pole terms (Appendix~\ref{app:S_pole}),
the FFR contributions (Appendix~\ref{app:FFR}), and in addition the loop and
counter terms. In order to compare with the dispersion integrals, we have to
subtract the pole terms. The results for the dispersive amplitudes take the
following form at lowest order:
\begin{eqnarray}
E_{0+,{\rm disp}}^{(p\pi^0){\rm thr}}&=& \frac{eg_{\pi N}M_\pi}{8\pi M_N^2}
\left\{\kappa_p(1-\frac{M_\pi}{M_N}) + \frac{M_NM_\pi}{16 \pi F_\pi^2}
\Xi_1\right\}\nonumber \\
&\approx& \left( 8.5 + 24.1\, \frac{Q^2}{\rm{GeV}^2} \right)\
\frac{10^{-3}}{M_{\pi^+}}\,,\label{eq:E0+neutral_disp}\\
L_{0+,{\rm disp}}^{(p\pi^0){\rm thr}}&=& E_{0+,{\rm disp}}^{(p\pi^0){\rm thr}}
+ \frac{eg_{\pi
N}}{16\pi\,M_N^3}(M_{\pi}^2 \, + Q^2) \nonumber \\
& \times & \left\{-\kappa_p +
\frac{M_N^2}{8 \pi F_\pi^2}\,\Xi_2\right\}\nonumber \\
&\approx& \left(6.9 - 48.7 \,\frac{Q^2}{\rm{GeV}^2} \right)\,,
\frac{10^{-3}}{M_{\pi^+}} \, , \label{eq:L0+neutral_disp}\\
E_{0+,{\rm disp}}^{(-){\rm thr}}&=&  \frac{eg_{\pi N}}{8\pi M_N^3}\bigg\{C_0
M_{\pi}^2 + \frac{M_N^2 M_{\pi}^2} {8 \pi^2 F_{\pi}^2} \Xi_3 \nonumber \\
& + & \frac {M_N^2 Q^2}{6}\bigg(\langle r^2 \rangle_1^V-\langle r^2 \rangle_A^V
\bigg)\bigg\}\nonumber \\
&\approx& \left( 0.8 + 19.4\, \frac{Q^2}{\rm{GeV}^2} \right)\
\frac{10^{-3}}{M_{\pi^+}}\, ,\label{eq:E0+minus_disp}\\
L_{0+,{\rm disp}}^{(-){\rm thr}}& = & E_{0+,{\rm disp}}^{(-){\rm thr}}+
\frac{eg_{\pi N}}{48\pi M_N}(M_{\pi}^2 \, + Q^2)\nonumber \\
& \times & \left\{ \langle r^2\rangle_A^V - \langle r^2\rangle_1^V  +\frac {3}
{4 \pi^2F_\pi^2}\,\Xi_4\right\} \nonumber \\
& \approx &  \left( 0.03 - 13.7 \,\frac{Q^2}{\rm{GeV}^2} \right)\,
\frac{10^{-3}}{M_{\pi^+}} \,. \label{eq:L0+minus_disp}
\end{eqnarray}
The numerical values in the above equations are obtained by a strict expansion
to ${\cal O} (q^2)$ with the radii and other constants as given in Sec.~5. In
order to judge the physical relevance of these leading order dispersive
effects, they have to be compared to the corresponding pole contributions given
by Appendix~\ref{app:S_pole}:
\begin{eqnarray}
E_{0+,{\rm pole}}^{(p\pi^0){\rm thr}} &=& \left(-7.9 + 56.8\,
\frac{Q^2}{\rm{GeV}^2} \right)\
\frac{10^{-3}}{M_{\pi^+}}\,,\label{eq:E0+neutral_pole}\\
L_{0+,{\rm pole}}^{(p\pi^0){\rm thr}} &=& \left(-7.9 +55.7
\,\frac{Q^2}{\rm{GeV}^2} \right)\,
\frac{10^{-3}}{M_{\pi^+}} \, , \label{eq:L0+neutral_pole}\\
E_{0+,{\rm pole}}^{(-){\rm thr}} &=& \left( 21.0 - 80.4\,
\frac{Q^2}{\rm{GeV}^2} \right)\
\frac{10^{-3}}{M_{\pi^+}}\, ,\label{eq:E0+minus_pole}\\
L_{0+,{\rm pole}}^{(-){\rm thr}} &=&  \left( 11.9-300.0
\,\frac{Q^2}{\rm{GeV}^2} \right)\, \frac{10^{-3}}{M_{\pi^+}} \,.
\label{eq:L0+minus_pole}
\end{eqnarray}
Let us first have a look at neutral pion electroproduction as described by
Eqs.~(\ref{eq:E0+neutral_disp}, \ref{eq:L0+neutral_disp}) and
(\ref{eq:E0+neutral_pole}, \ref{eq:L0+neutral_pole}). It is seen that the
dispersive and pole contributions cancel to about one order of magnitude at
$Q^2=0$, which leads to very small physical photoproduction amplitudes at
threshold. Also the slopes in $Q^2$ take large absolute values for both
contributions. However, they add for the electric amplitude, whereas there is
again a large cancelation for the longitudinal S-wave amplitude. Therefore,
$L_{0+}^{(p\pi^0){\rm thr}}$ is significantly suppressed relative to the
transverse S-wave multipole. Because the pole amplitude is well defined, we may
conclude that the physically interesting dispersive amplitude can be well
determined for both multipoles, although the respective cross sections are much
smaller than expected from the size of the pole terms. The situation is quite
different for the isospin $(-)$ multipoles to be measured by charged pion
production. Equations~(\ref{eq:E0+minus_disp}, \ref{eq:L0+minus_disp}) show
that the dispersive effects are small compared to the pole contributions,
Eqs.~(\ref{eq:E0+minus_pole}, \ref{eq:L0+minus_pole}). In particular in the
real photon limit, the dispersive contributions are at most a few per cent of
the respective pole contributions. The same is true for the slope of the
longitudinal S wave, which has an extremely large pole contribution. In
conclusion, the corresponding cross sections are large but the physically
interesting dispersive contributions are hidden under the large ``background''
of the pole terms. The only exception is the slope of $E_{0+,{\rm
disp}}^{(-){\rm thr}}$ containing information on $\langle r^2\rangle_1^V-
\langle r^2\rangle_A^V$, which takes the value 0.141~fm$^2$ with the radii
given in Sect.~5. However, as has been pointed out by Bernard {\emph et
al.}~\cite{Ber92c}, the loop function $\Xi_3$ yields a sizeable loop correction
of 0.046~fm$^2$. It is therefore mandatory to include the loop effects in the
data analysis for pion electroproduction. One should also keep in mind that the
experiment is dominated by the pole term of Eq.~(\ref{eq:E0+minus_pole}), that
is, the dispersive contribution yields only about $10~\%$ of the total
threshold multipole $E_{0+}^{(-)}$ at $Q^2$=0.05~GeV$^2$. In the real photon
limit, the information on the radii is contained in the longitudinal S-wave
multipole, $L_{0+,{\rm pole}}^{(-){\rm thr}}$. The numerical value of the
dispersive contribution, Eq.~(\ref{eq:L0+minus_disp}), is however dwarfed by
the pole term, Eq.~(\ref{eq:L0+minus_pole}). We conclude that the longitudinal
S-wave multipole has no practical relevance for studies of the axial radius,
because the radius-dependent term and the loop corrections have different signs
resulting in a very small net effect, the remaining small value is likely to
change by higher-order loop effects, and the pole contribution is larger than
the predicted dispersive effect by about two orders of magnitude. After this
review of the key phenomena as predicted by ChPT, we present our dispersive
results in the following section.
\section{Results and Discussion}
The numerical calculations in this section are performed with the following
values: $g_{\pi N}=13.4$, $e$=$\sqrt{4 \pi/137}$, $M_N=0.938$~GeV,
$M_{\pi^+}=0.140$~GeV, $M_{\pi^0}=0.135$~GeV, $F_{\pi}=0.0924$~GeV,
$\kappa_p=1.793$, and $\kappa_n=-1.913$.
\subsection{Dispersive contributions to the invariant amplitudes}
\begin{figure*}[htb]
\begin{center}
\epsfig{figure=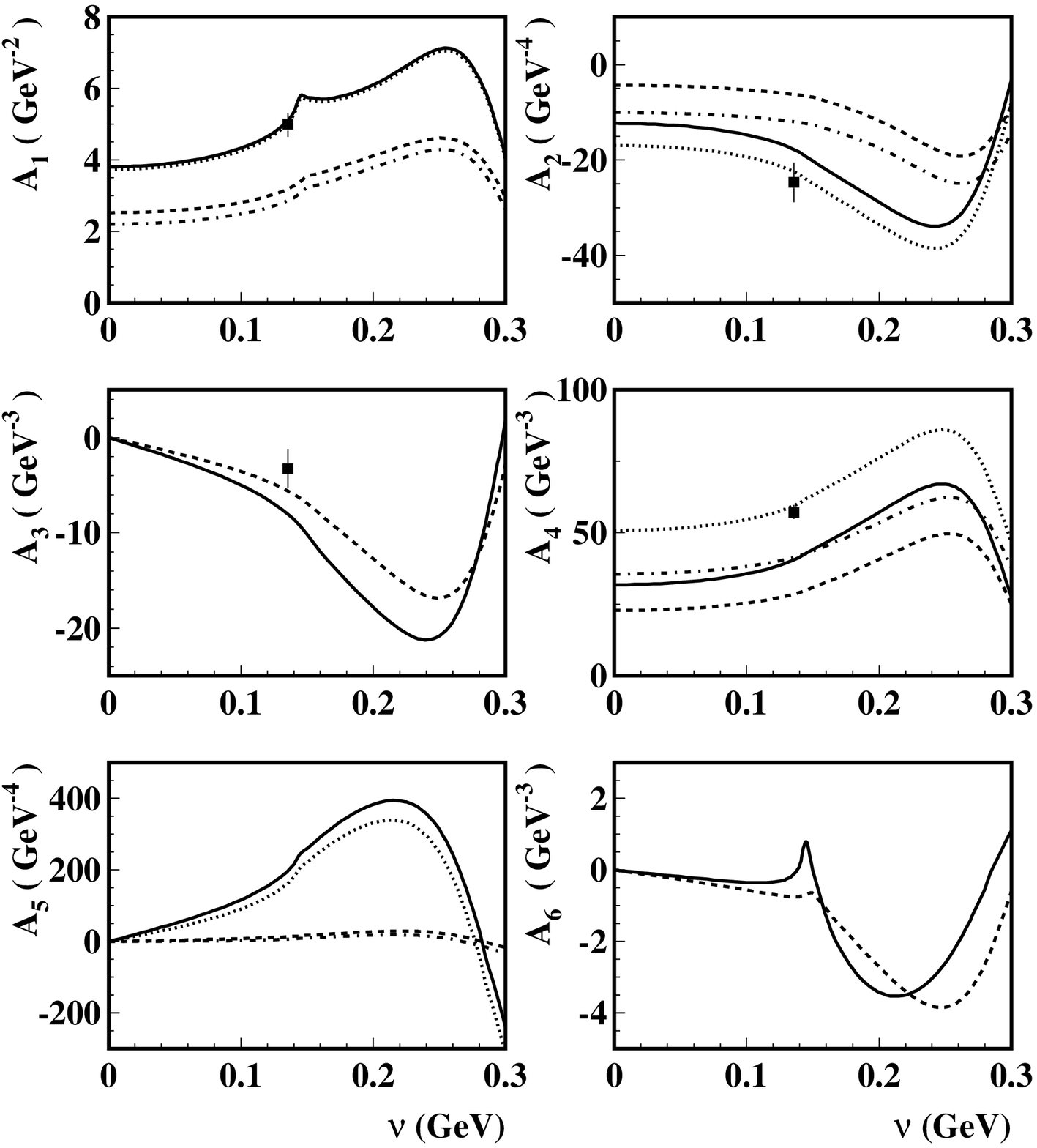,width=8.7cm,angle=0}
\epsfig{figure=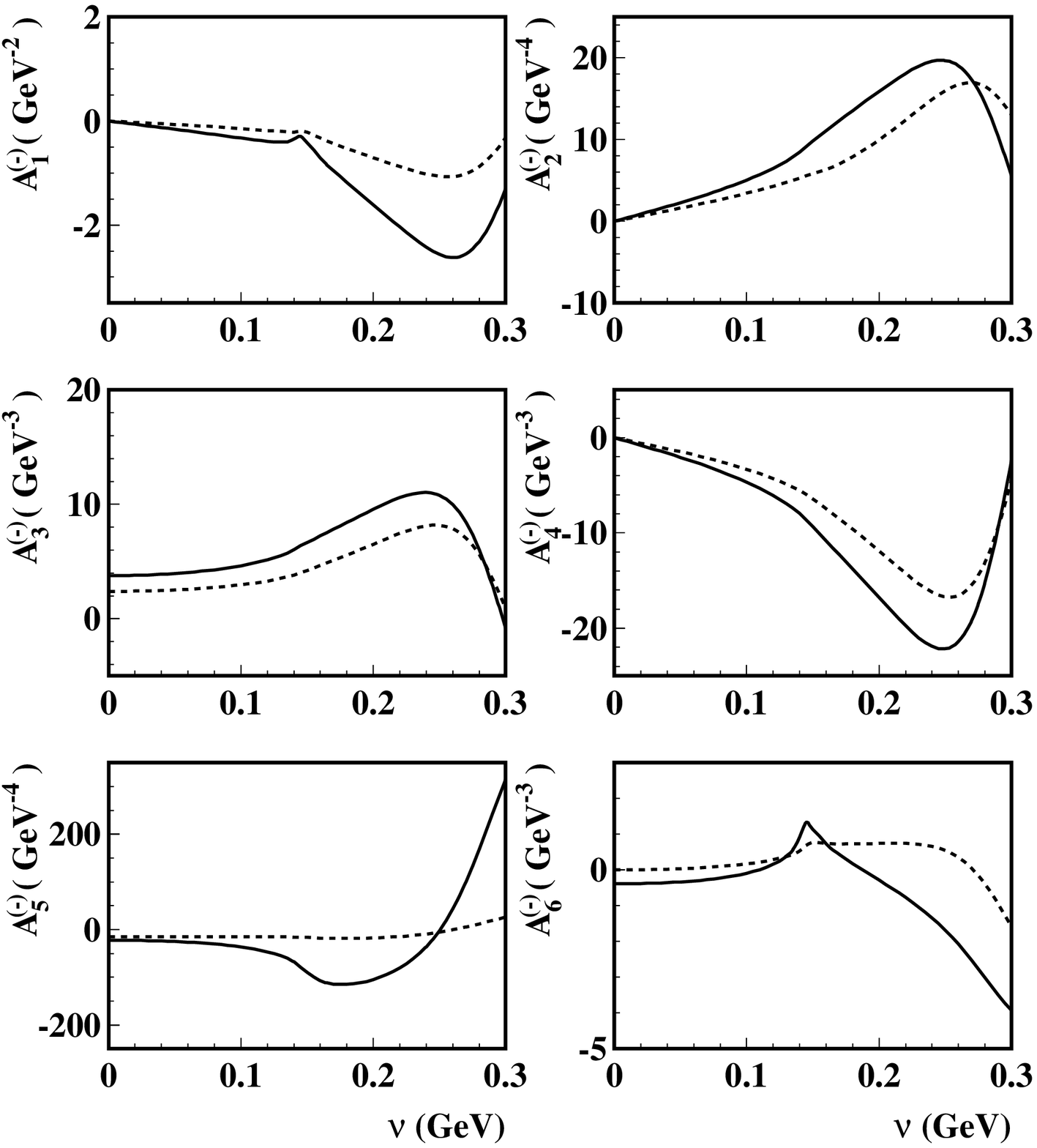,width=8.7cm,angle=0}
\end{center}
\caption{The invariant amplitudes for pion electroproduction at $t=t_{\rm
thr}(Q^2)$ as function of $\nu$. Solid lines: dispersive contributions for
$Q^2=0$, dashed lines: same at $Q^2=0.1$~GeV$^2$. In the p$\pi^0$ channel (left
panels), the inclusion of the vector meson poles leads to the dotted ($Q^2=0$)
and dashed-dotted lines ($Q^2=0.1$~GeV$^2$). The data points for real photons
are derived from the experimental values of Ref.~\cite{Sch01}. The isospin
$(-)$ channel is shown in the right panels.} \label{fig:Ai_nu}
\end{figure*}

\begin{figure*}[htb]
\begin{center}
\epsfig{figure=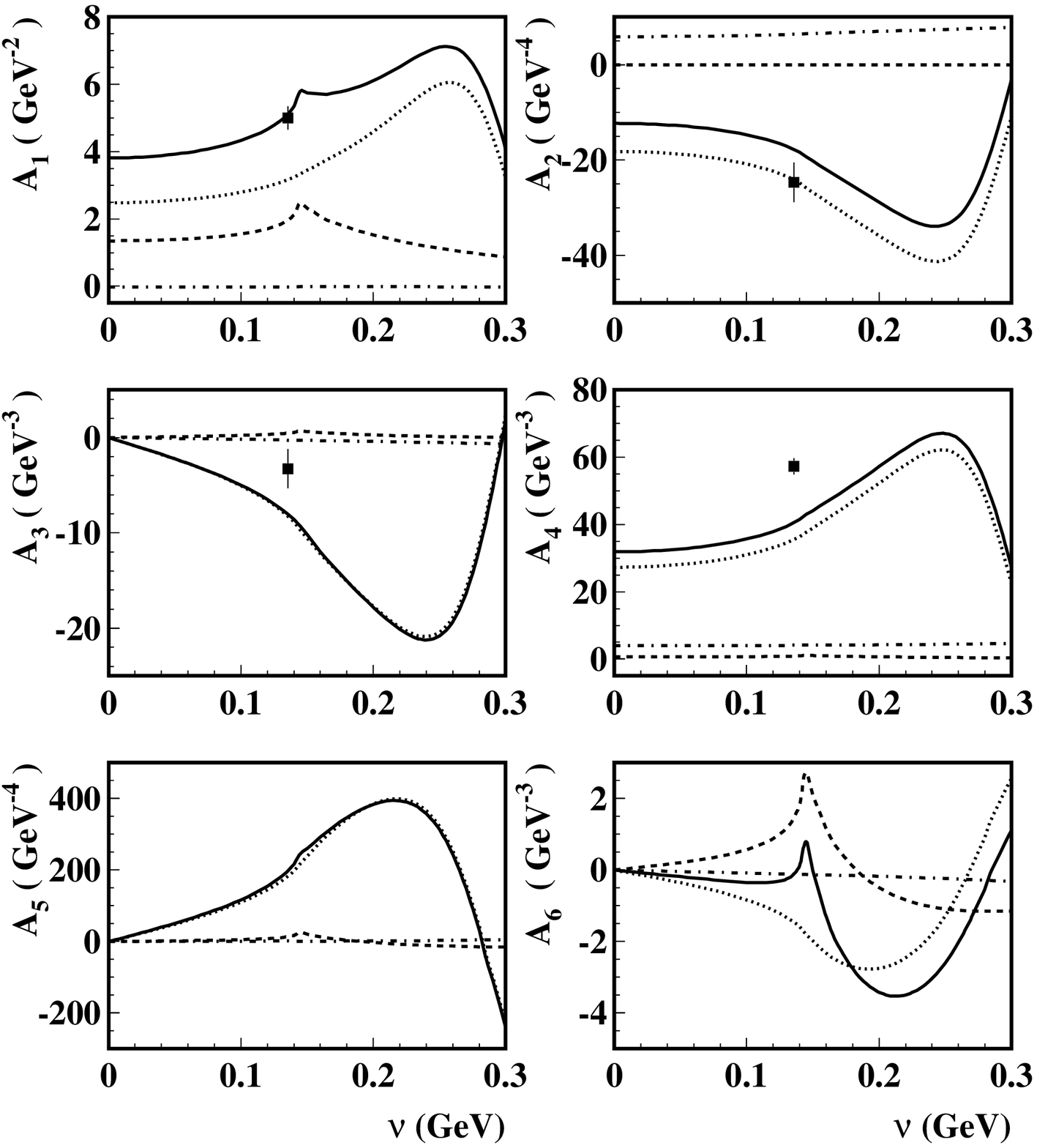,width=8.7cm,angle=0}
\epsfig{figure=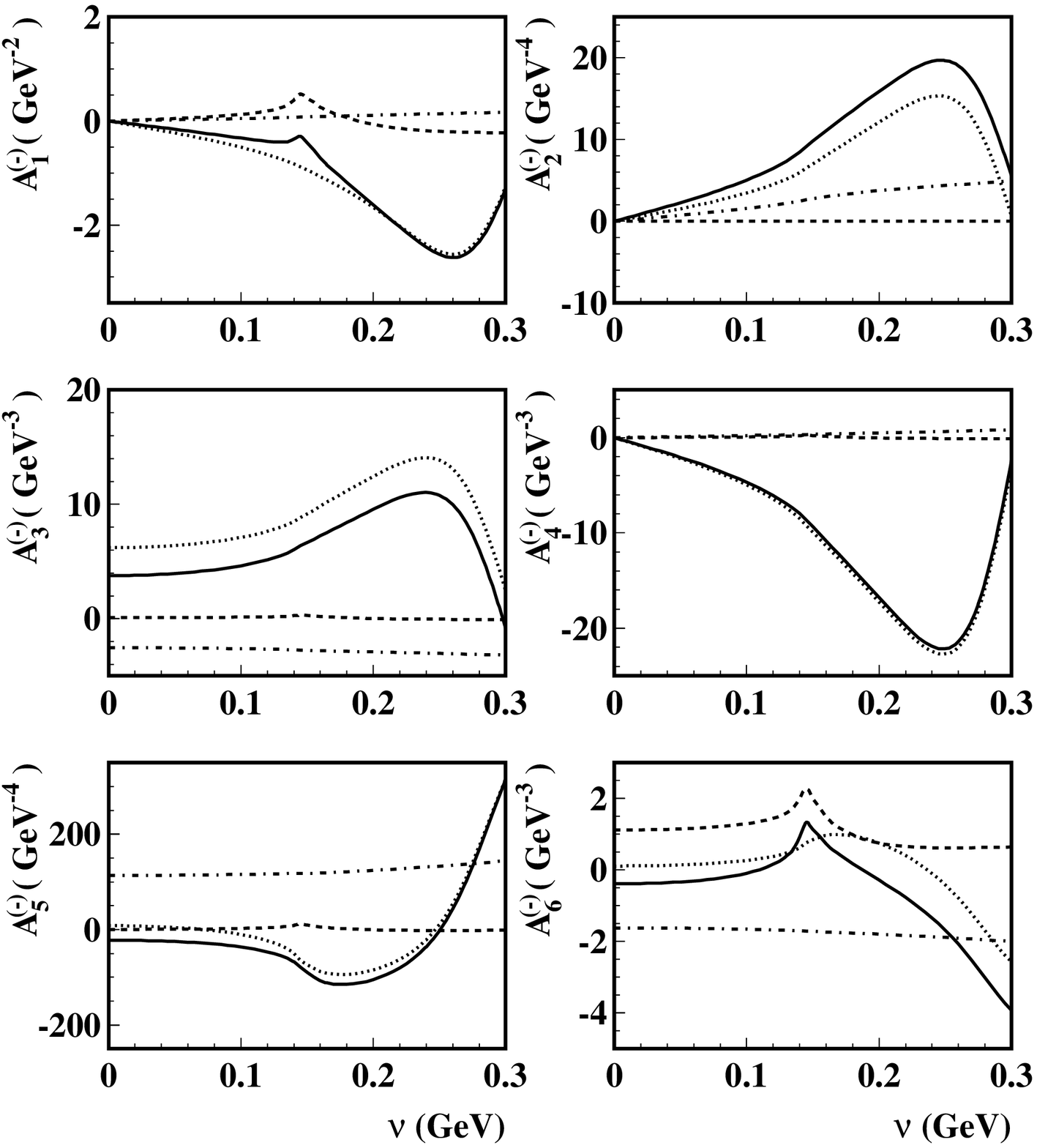,width=8.7cm,angle=0}
\end{center}
\caption{Contributions to the invariant amplitudes from the imaginary part of S
waves (dashed lines), P waves (dotted lines), D plus F waves (dashed-dotted
lines), and total result (solid lines). See Fig.~\ref{fig:Ai_nu} for further
notation.} \label{fig:integrand_mult}
\end{figure*}

The dispersive contributions to the invariant amplitudes are shown as function
of the crossing-symmetric variable $\nu$ in Fig.~\ref{fig:Ai_nu} for the
p$\pi^0$ and isospin $(-)$ channels, respectively. Even though all calculations
are isospin symmetric, we define the threshold as $W=M_p+M_{\pi^0}$ for the
isospin (+,0) channels and $W=M_n+M_{\pi^+}$ for the $(-)$ channel. The Wigner
cusp at the onset of charged pion production is clearly seen for the amplitudes
$A_1$ and $A_6$, which receive large contributions from the multipoles $E_{0+}$
and $L_{0+}$. Comparing the results for real photons (solid lines, $Q^2=0$) and
virtual photons (dashed lines, $Q^2=0.1$~GeV$^2$), we find strong differences
for some of the amplitudes over this moderate range of the virtuality. As has
been shown in our previous work~\cite{Pas05,Pas06}, the pole contributions due
to the $t$-channel exchange of vector mesons must be included explicitly (see
the dotted lines) in order to agree with the data for neutral pion production.
For charged pion production, it is not possible to construct the dispersive
amplitudes directly from the threshold data in a reliable way, because the pole
contributions dominate this region. The contributions to the dispersion
integral from the S, P, and higher multipoles are shown in
Fig.~\ref{fig:integrand_mult}. It is seen that the S-wave contribution is
strong for the real part of the threshold amplitudes $A_1$ and $A_6$, whereas
all the other amplitudes are dominated by the P waves in the imaginary part of
the amplitudes. The following Fig.~\ref{fig:Ai_Q2} displays the dispersive
contributions to the invariant amplitudes as function of the virtuality  $Q^2$.
In most cases the amplitudes change rather dramatically over the range $0 \leq
Q^2 \leq 0.1~{\rm GeV}^2$.
\begin{figure*}[htb]
\begin{center}
\epsfig{figure=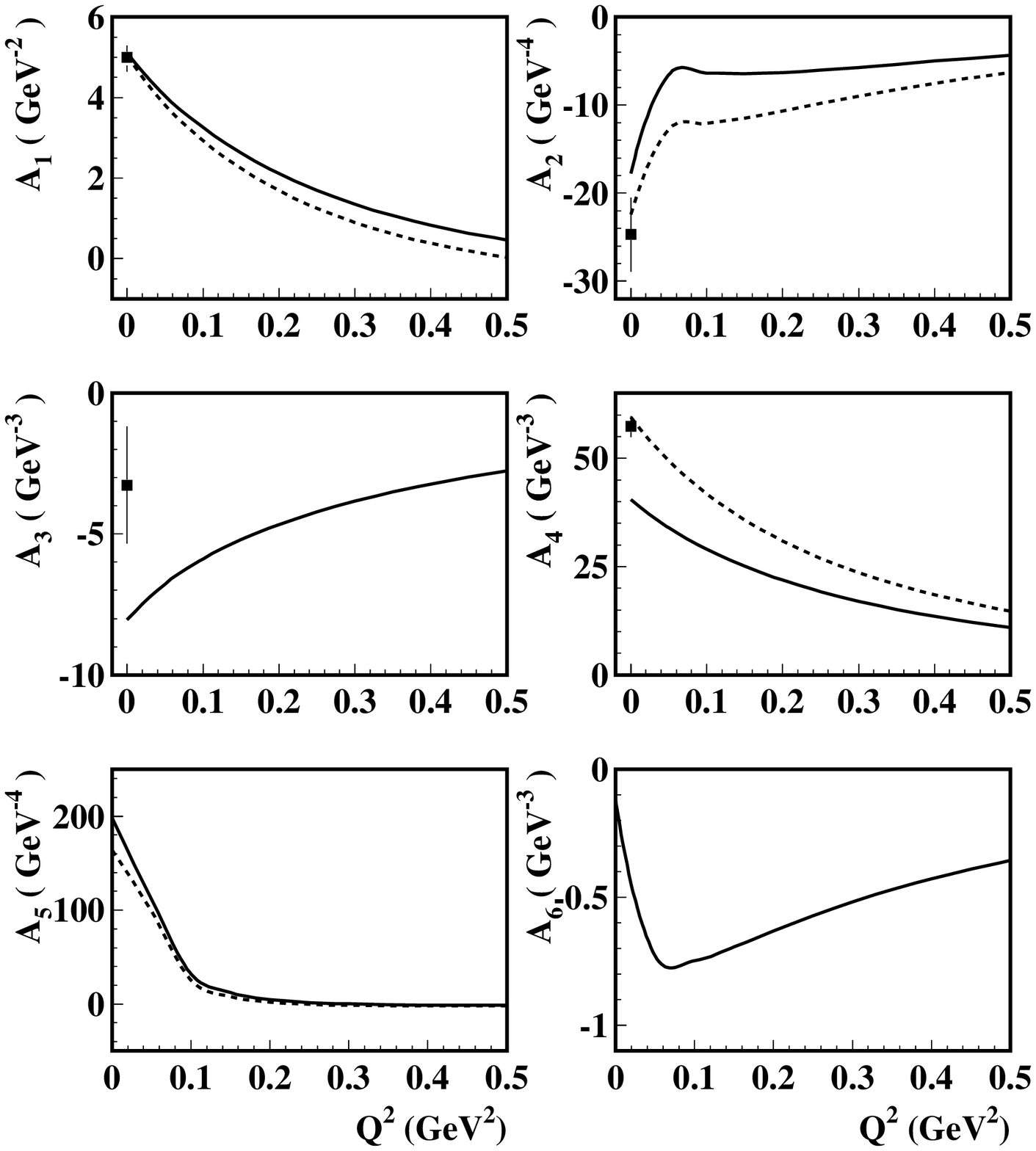,width=8.7cm,angle=0}
\epsfig{figure=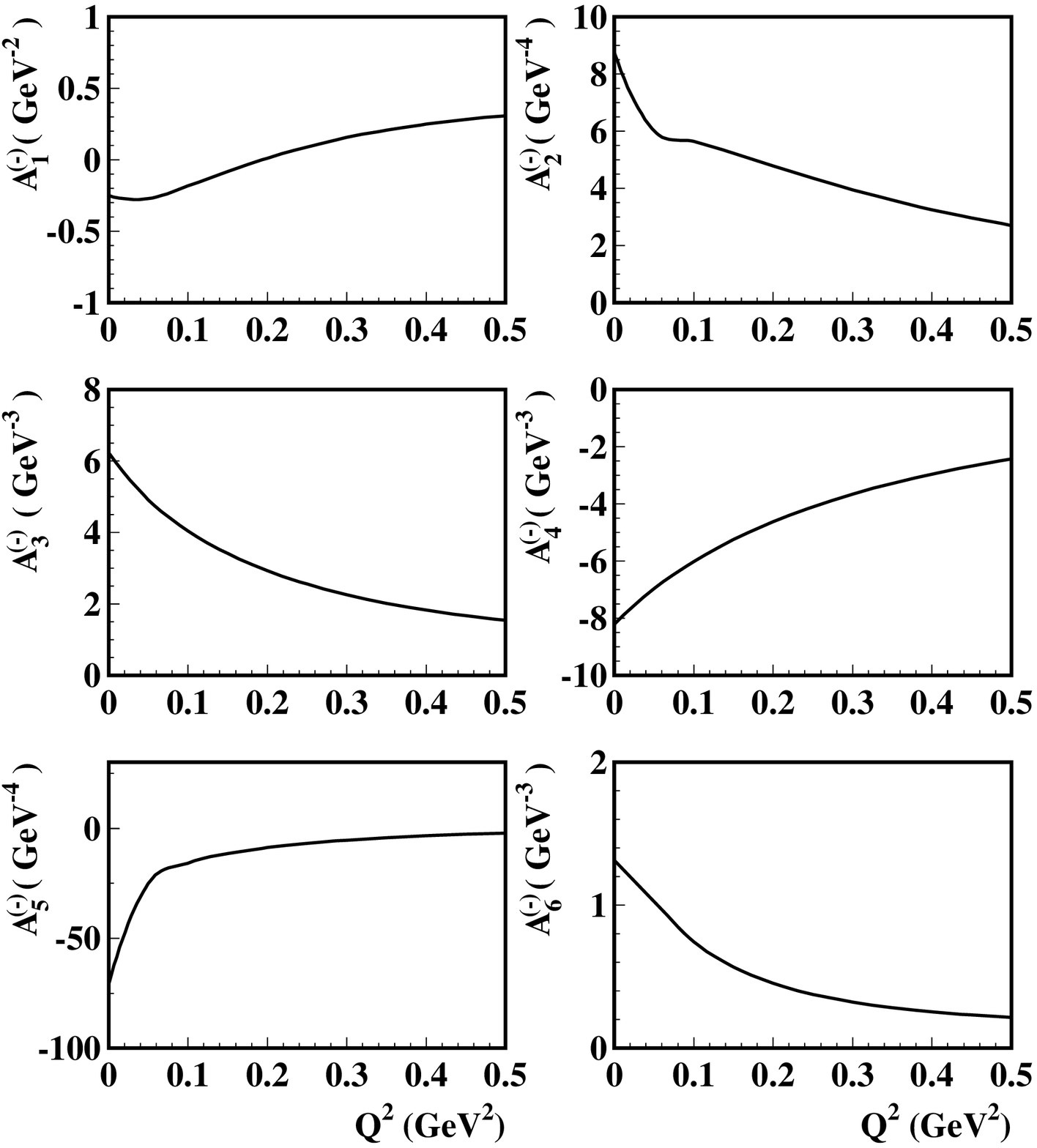,width=8.7cm,angle=0}
\end{center}
\caption{The invariant amplitudes for pion electroproduction at
$t=t_{\rm thr}(Q^2)$ and $\nu=\nu_{\rm thr}(Q^2)$ as function of
$Q^2$. For the $p\pi^0$ channel (left panel) the inclusion of the
vector meson poles leads to the dashed curves.
}
\label{fig:Ai_Q2}
\end{figure*}
\subsection{The FFR sum rule for the invariant amplitude $A_1$}
For small values of the variables, the dispersive part of the
invariant amplitudes can be expanded in a Taylor series. As an
example, we cast Eq.~(\ref{eq:FFRQ2}) into the form
\begin{equation}\label{eq:A1_disp}
A_{1,{\rm disp}}^{(p \pi^0)}(\nu,\nu_B,Q^2) =  \frac{eg_{\pi N}}{2M_N^2}
\,\left(F_2^p(Q^2) +\Delta(\nu,\nu_B,Q^2)\right),
\end{equation}
with $\Delta \equiv \Delta_1^{(p \pi^0)}$ the dimensionless ``FFR
discrepancy''. This function is real in a triangle defined by the straight
lines $s=(M_N+M_{\pi})^2$, $u=(M_N+M_{\pi})^2$, and $t=4 M_{\pi}^2$, which
define the onset of particle production in the $s$, $u$, and $t$ channels,
respectively. In this region of the Mandelstam plane, the crossing-even
function $\Delta_1^{(p \pi^0)}$ has the expansion
\begin{equation}\label{eq:Delta1_disp}
\Delta(\nu,\nu_B,Q^2)  =  \delta_0 + \delta_{\nu}\frac{\nu^2} {M_{\pi}^2} +
\delta_B\frac {\nu_B} {M_{\pi}} + \delta_Q Q^2/M_\pi^2 + ...
\end{equation}
The dispersive amplitude in Eq.~(\ref{eq:A1_disp}) is evaluated by the
dispersion integral at $t = t_{\rm thr}(Q^2)$ along the path from $\nu =
\nu_{\rm thr} (Q^2)$ to infinity. In the soft-pion kinematics, the threshold
moves to $\nu=0$ and $\nu_B = 0$ (or $t=M_{\pi}^2-Q^2$ as long as the pion mass
is finite). For small values of $\nu$ and $\nu_B$ we can use
Eq.~(\ref{eq:Delta1_disp}) to extrapolate from the physical to the soft-pion
threshold. Of course, we can not expect to reproduce the FFR sum rule in this
way, because the expansion coefficients in Eq.~(\ref{eq:Delta1_disp}) depend on
the pion mass and the dispersion calculation only provides these coefficients
for the physical mass. In particular the pion loop effects at threshold depend
on the pion mass and, moreover, produce a $Q^2$ dependence very different from
the nucleon form factors. However, from previous experience~\cite{Pas05,Pas06}
we might expect a suppression of these loop effects if the dispersion integral
is evaluated in the sub-threshold region.
\begin{figure}[ht]
\begin{center}
\epsfig{figure=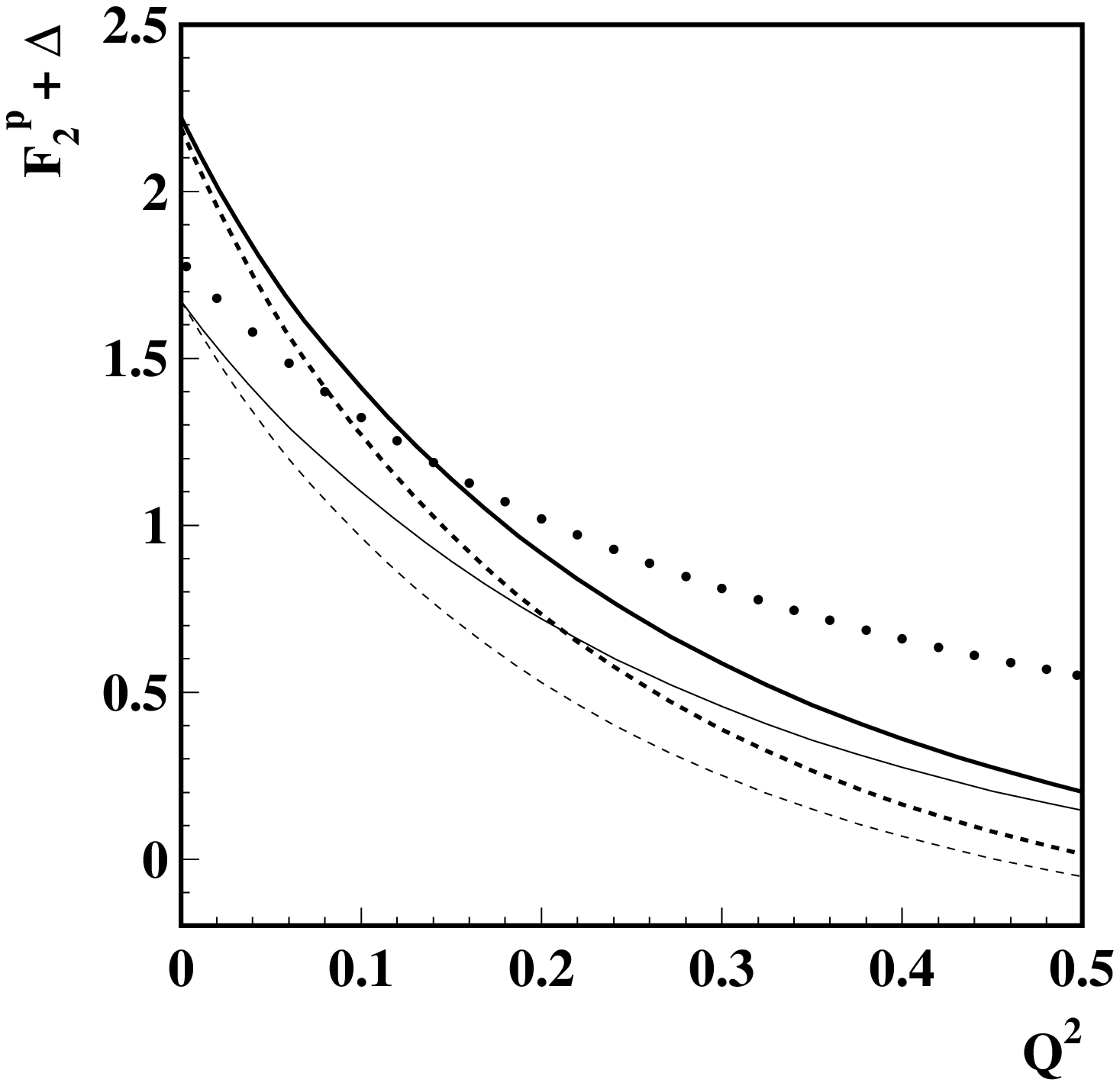,width=8cm,angle=0}
\end{center} \caption{The proton's Pauli form factor $F_2^p$ as function of
$Q^2$, compared to the result of dispersion theory as obtained from the rhs of
Eq.~(\ref{eq:FFRQ2}). Thick solid line: dispersive results for the threshold
amplitude ($\nu=\nu_{\rm thr}(Q^2), t=t_{\rm thr}(Q^2)$), thick dashed line:
same kinematics but including the $t$-channel vector meson poles, thin solid
line: dispersive results for soft-pion kinematics ($\nu=\nu_B=0$), thin dashed
line: same kinematics but including the $t$-channel vector meson poles. The
dashed-dotted line is the parametrization of $F_2^p$ according to
Ref.~\cite{Kel04}.} \label{fig:A1p&FFR}
\end{figure}
In Fig.~\ref{fig:A1p&FFR} we compare the Pauli form factor $F_2^p(Q^2)$~(dotted
line) to the $Q^2$ dependence of $A_{1, \,{\rm disp}}^{(p \pi^0)}$ as evaluated
by the dispersion integral at $\nu=0$ and $\nu = \nu_{\rm thr}(Q^2)$. Whereas
the deviations from the sum rule are quite sizeable at physical threshold
(thick lines), they indeed decrease if $\nu$ moves towards 0. Only slight
changes occur if we further extrapolate from $t=t_{\rm thr}(Q^2)$ to the
soft-pion kinematics (thin lines) at $\nu=\nu_B=0$. However,
Fig.~\ref{fig:A1p&FFR} clearly demonstrates that the slopes of the Pauli form
factor and the invariant amplitude $A_{1,\,{\rm disp}}^{(p \pi^0)}$ differ
quite a bit. Even the extrapolation of the invariant amplitude to the soft-pion
kinematics yields an effective r.m.s. radius much larger than the Pauli radius
of the proton, $r_2^p = 0.894$~fm of Ref.~\cite{Mer96} or $r_2^p = 0.879$~fm of
Ref.~\cite{Kel04}.
\begin{figure}[htb]
\begin{center}
\epsfig{figure=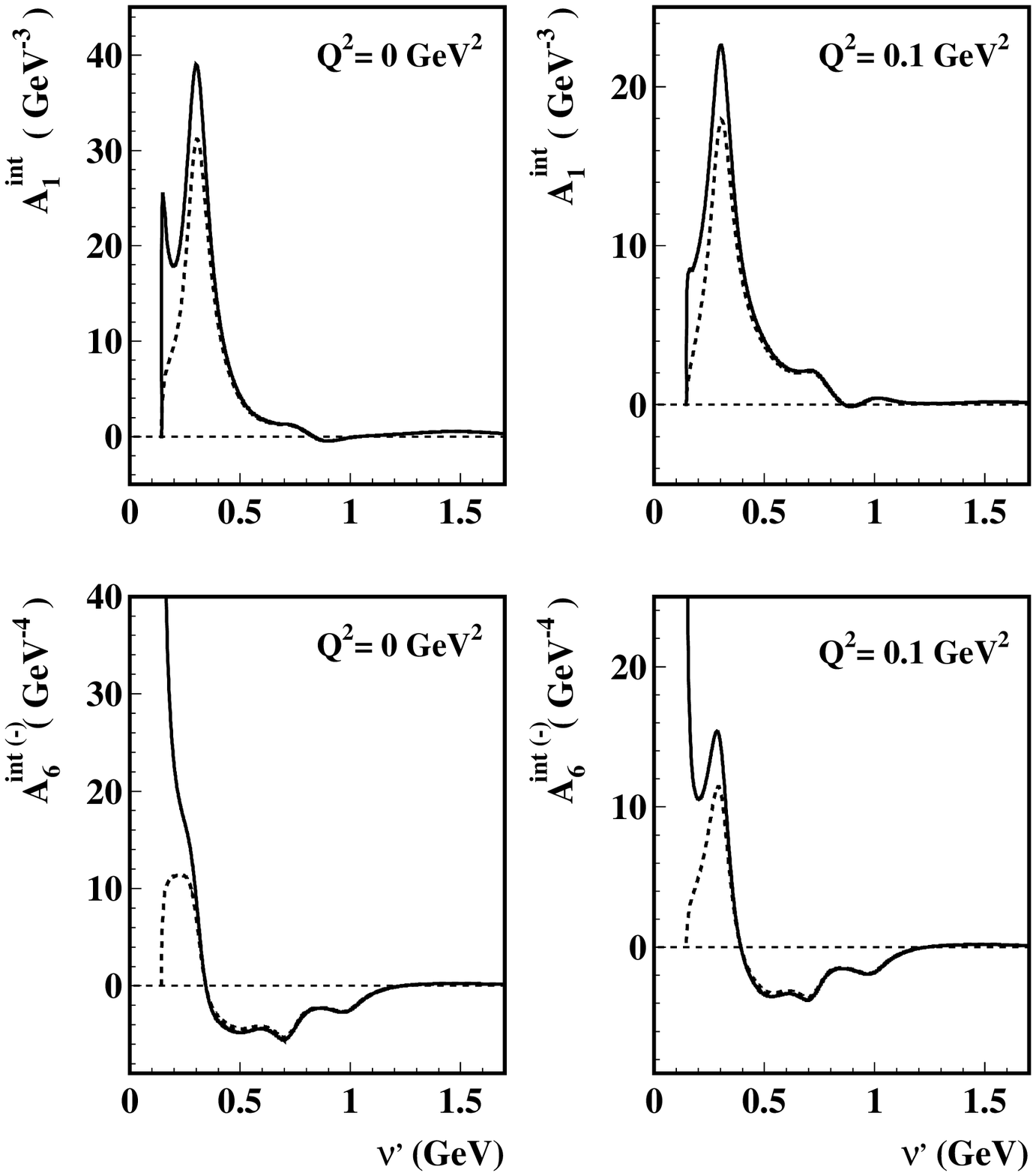,width=9cm,angle=0}
\end{center}
\caption{The integrands of the dispersion integrals for $A_1^{(p \pi^0)}$ and
$A_6^{(-)}$ are shown in the upper and lower rows, respectively. Left: $Q^2=0$,
right: $Q^2=0.1$~GeV$^2$; solid lines: integrands obtained for $\nu= \nu_{\rm
thr}(Q^2)$, dashed lines: integrands obtained for $\nu=0$.}
\label{fig:integrand}
\end{figure}
Let us now study the physics behind these deviations by looking at the
integrand and the multipole decomposition of the dispersion integral. The
integrand for the amplitude $A_{1,\,{\rm disp}}^{(p \pi^0)}$ is shown in the
top panels of Fig.~\ref{fig:integrand} for the momentum transfers $Q^2=0$ and
$Q^2=0.1$~GeV$^2$. Evidently the bulk contribution to the integral stems from
the $\Delta(1232)$ resonance. In the real photon limit and for energies near
threshold (solid line) also the S-wave threshold production is quite sizeable,
but this contribution of the pion cloud decreases rapidly if the energy moves
into the sub-threshold region (dashed line). It is also seen that the loop
effects drop faster with momentum transfer $Q^2$ than the resonance
contributions. The contributions of the most important multipoles at $Q^2=0$
are, in units of GeV$^{-2}$,
\begin{eqnarray}
&&A_{1,\,{\rm disp}}^{(p\pi^0)\,{\rm thr}}(0) = 1.97~(E_{0+}) + 3.59~(M_{1+})
-0.41~(M_{1-})\nonumber\\
&& \quad  +0.10~(E_{2-}) - 0.30~(M_{2-}) + 0.17~({\rm{others}}) \nonumber\\
&&\quad = 5.12. \label{eq:multA1}
\end{eqnarray}
This has to be compared with the sum rule value given by Eq.~(\ref{eq:A1FFR}),
\begin{equation}
\label{eq:A_1_LET_alt} A_{1, {\rm FFR}}(Q^2) = (4.14 -13.68 \frac {Q^2}{{\rm
GeV}^2}+ \ldots)~{\rm GeV}^{-2}\,.
\end{equation}
The difference between the dispersive calculation and the FFR
prediction demonstrates the importance of the pion loops near
threshold. If the integral is evaluated at $\nu = 0$, the S-wave
contribution decreases to $1.35$~GeV$^{-2}$ and the total result is
$A_{1,\,{\rm disp}}^{(p\pi^0)}(\nu = 0, t_{\rm thr},Q^2=0)=
3.81$~GeV$^{-2}$. Finally, the extrapolation to the soft-pion
kinematics leads to $A_{1,\,{\rm disp}}^{(p\pi^0)}(\nu = \nu_B =
Q^2=0)= 3.90$~GeV$^{-2}$, quite close to the sum rule. However, as
shown in Fig.~\ref{fig:A1p&FFR}, the slope of this function differs
from the FFR prediction even in the sub-threshold region ($\nu
\rightarrow 0$). The reason for this behavior is already seen in a
simple model including the loop contributions in the S waves, see
Eqs.~(\ref{eq:E0+neutral_disp}) and (\ref{eq:L0+neutral_disp}), plus
the FFR contributions for all the multipoles (see
Appendix~\ref{app:FFR}), both in the unexpanded form. Within this
model we obtain the following slope for the threshold amplitude and
its contributions, all in units of GeV$^{-4}$:
\begin{eqnarray}
&&\frac {d}{d\, Q^2}\,A_{1,\,{\rm disp}}^{(p\pi^0)\,{\rm
thr}}({\rm{model}}) \nonumber\\
&&\quad =-311~(E_{0+}) + 278~(L_{0+})-1~({\rm{P~waves}})\nonumber \\
&&\quad = -13.5~({\rm{FFR}}) - 20.5~({\rm{loop}}) = -34. \label{eq:modelA1}
\end{eqnarray}
The strong cancelation of the transverse and longitudinal S-wave
contributions is remarkable. We conclude that a precise knowledge of
both multipoles is required in order to get a reliable prediction
for the slope. Furthermore, the slope receives large loop
contributions. Translated into transition radii, the FFR term has
the radius of the Pauli form factor, $r_2^p \approx 0.88$~fm,
whereas the pion cloud reaches to a much larger distance described
by $r[{\rm {loop}}]\approx 1/M_{\pi} = 1.45$~fm. The total result is
$r[{\rm {model}}] = 1.12$~fm, in good agreement with the following
results obtained from the dispersion integral: $r[A_{1,\,{\rm
disp}}^{(p\pi^0)}(\nu_{\rm thr},\, t_{\rm thr},0)] = 1.16$~fm,
$r[A_{1,\,{\rm disp}}^{(p\pi^0)}(0,\, t_{\rm thr},0)] = 1.08$~fm,
and $r[A_{1,\,{\rm disp}}^{(p\pi^0)}(0,\,M_{\pi}^2,0)] = 1.13$~fm.
In conclusion, the FFR sum rule can not be used to determine the
Pauli form factor from the $Q^2$ dependence of the invariant
amplitude $A_1$. The radius derived from that observable is about
25~\% larger than the Pauli radius, which is another ``smoking gun''
for the importance of the pion cloud in low-energy nuclear physics.
\subsection{The FFR sum rule for the invariant amplitudes $A_6$}
\begin{figure*}[htb]
\begin{center}
\epsfig{figure=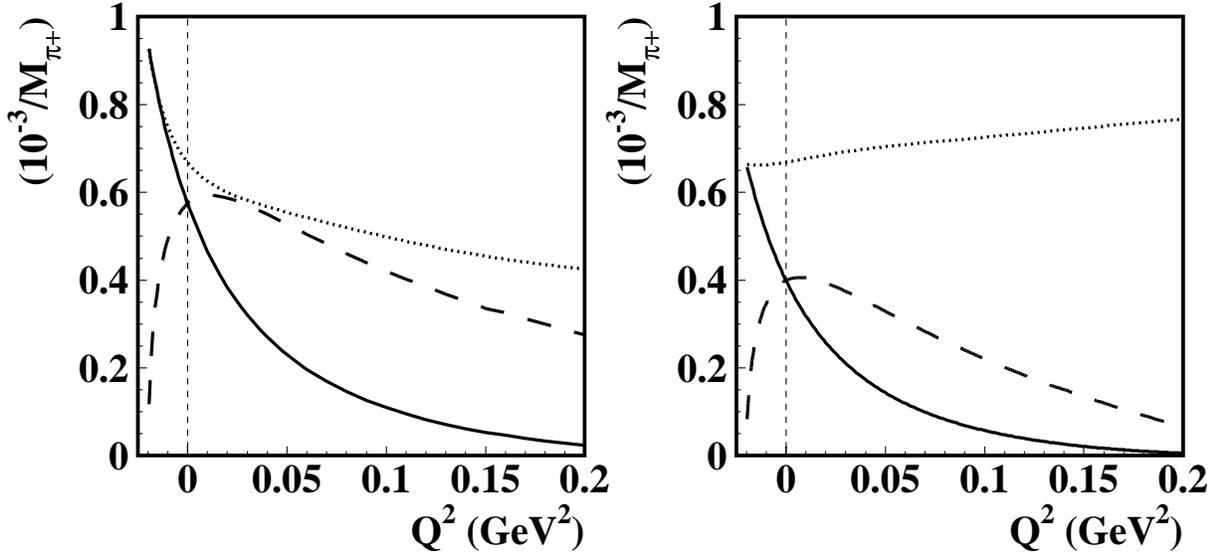,width=16cm,angle=0}
\end{center}
\caption{The S-wave multipoles $E_{0+}^{(-)}$ (dotted lines), $L_{0+}^{(-)}$
(solid lines), and $S_{0+}^{(-)}$ (dashed lines) as function of $Q^2$ at $\nu =
\nu_{\rm {thr}}(Q^2)$ and $t = t_{\rm {thr}}(Q^2)$, in units of
$10^{-3}/M_{\pi^+}$. Left panel: results with MAID05 as input for the
dispersion integral, right panel: same with MAID07~\cite{MAID07}. Since the
dispersive contributions are only a small fraction of the full multipoles, the
shown deviations between the two version of MAID are within the present
experimental error bars.} \label{fig:S_waves}
\end{figure*}
Let us now turn to Eq.~(\ref{eq:A6FFR}), which connects the axial and Dirac
isovector form factors with the amplitude $A_6^{(-)}(\nu, t_{\rm thr}(Q^2),
Q^2)$. The isovector Dirac radius is relatively well known from various
analyses of elastic electron scattering, e.g., $\langle r^2\rangle_1^V =
(0.585\pm0.010)$~fm$^2$~\cite{Mer96}. The axial mass parameter as determined by
neutrino and antineutrino scattering~\cite{Ahr88} lies in the range of
$M_A=(1.026\pm0.021)$~GeV corresponding to $\langle r^2\rangle_A^V =
(0.444\pm0.019)$~fm$^2$~\cite{Lie99}. With these values we obtain
\begin{equation}
\label{eq:A_6_LET} A_{6, {\rm FFR}}(0)=(1.31\pm0.27)~{\rm GeV}^{-3}\, ,
\end{equation}
and with the same axial mass but the form factor parametrization of
Kelly~\cite{Kel04}
\begin{equation}
\label{eq:A_6_LET_alt} A_{6, {\rm FFR}}(Q^2) = (1.54 - 9.48\frac {Q^2}{{\rm
GeV}^2}+ \ldots)~{\rm GeV}^{-3}\,.
\end{equation}
According to Eq.~(\ref{eq:thresh_A6}) of Appendix~\ref{app:A5A6},
only the S-wave multipole $E_{0+}^{(-)}(Q^2)$ survives in the
soft-pion limit $\mu\to 0$ as long as $Q^2$ is finite. In accordance
with the LET of Nambu {\it et al.}~\cite{Nam62}, the information of
the LET resides in the slope of that multipole. On the other hand,
the FFR current is purely longitudinal for $Q^2=0$ and finite pion
mass (see Appendix~\ref{app:FFR}). For the form factor
parametrization of Kelly~\cite{Kel04}, the multipole $L_{0+}^{(-)}$
accounts for 94~\% of $A_{6,\,\rm{FFR}}$ at $Q^2=0$, the remainder
being given by $L_{1-}^{(-)}$. Already at $Q^2=0.1$~GeV$^2$, the
bulk contribution (78\%) is due to the rising multipole
$E_{0+}^{(-)}$. In the real world of finite pion masses, the
situation is more complicated. The integrand for the amplitude
$A_{6,\, {\rm disp}}^{(-)}$ is shown in the lower panels of
Fig.~\ref{fig:integrand} for the momentum transfers $Q^2=0$ and
$Q^2=0.1$~GeV$^2$. The figure shows positive contributions from both
threshold pion production and $\Delta(1232)$ resonance excitation.
However, these contributions are largely canceled by equally strong
ones with opposite signs in the second and third resonance regions.
In order to quantify this effect, we decompose the imaginary part of
the amplitude into a multipole series. In this way the dispersive
photoproduction amplitude at the cusp takes the following form:
\begin{eqnarray}
&&A_{6,\,{\rm disp}}^{(-)\,{\rm cusp}}(0) = 3.82~(E_{0+}) -1.49~(L_{0+}) +
0.91~(M_{1+})\nonumber\\
&& + 1.05~(E_{1+})- 0.95~(L_{1+})+0.05~(M_{1-})-0.38~(L_{1+})\nonumber\\
&& - 1.80~(E_{2-})+0.06~(M_{2-})+0.38~(L_{2-})- 0.33~({\rm{rest}})\nonumber\\
&& =2.33~({\rm{S}})+0.68~({\rm{P}})-1.34({\rm{D}})-0.35~({\rm{F}})= 1.32,
\label{eq:multA6}
\end{eqnarray}
with all the values given in units of GeV$^{-3}$. The dispersive amplitude at
the cusp, Eq.~(\ref{eq:multA6}), confirms the FFR sum rule value given by
Eq.~(\ref{eq:A_6_LET}). This is surprising because of the formidable
cancelations occurring in Eq.~(\ref{eq:multA6}). We observe a substantial
cancelation both among multipoles with the same pion partial wave and between
the strong electromagnetic dipole excitations $E_{0+}^{(-)}$ and
$E_{2-}^{(-)}$. It is also remarkable that the electric transverse and
longitudinal multipoles of the $\Delta$~(1232) resonance, $E_{1+}^{(-)}$ and
$L_{1+}^{(-)}$, contribute just as much as the magnetic $M_{1+}^{(-)}$
transition, although the latter multipole is stronger by factors of 40 and 25,
respectively. Furthermore, both S and P waves yield positive contributions,
whereas the D and F waves of the second and third resonance regions diminish
the integral. As $\nu$ moves from the cusp value to $\nu=0$, the total S-wave
contribution decreases from 2.33~GeV$^{-3}$ to 1.12~GeV$^{-3}$ and the
$\Delta$(1232) contribution drops from 1.01~GeV$^{-3}$ to 0.59~GeV$^{-3}$,
whereas the higher multipole contributions change little. As a result the
invariant amplitude becomes negative, $A_{6,\,{\rm disp}}^{(-)}(\nu=0, t_{\rm
thr}, 0)= -0.38~{\rm {GeV}}^{-3}$. Incidentally, the discussed model of loop
and FFR terms yields a contribution of 4.05~GeV$^{-3}$ from the real part of
the threshold multipole $E_{0+}^{(-)}$, in qualitative agreement with
Eq.~(\ref{eq:multA6}). However, the multipole contributions of
Eq.~(\ref{eq:multA6}) are defined by a decomposition of the imaginary part, and
a particular multipole in the imaginary part will generally contribute to all
the multipoles in the real part. Furthermore, the large cancelation due to
higher resonances would have to be described by appropriate low-energy
constants in an effective field theory.\\

In spite of the modern precision data serving as input for the dispersive
calculation, the results still keep changing. This is demonstrated by
Fig.~\ref{fig:S_waves} showing the S-wave multipoles at threshold as function
of $Q^2$. The left panel displays these multipoles with input from MAID05, the
right panel is obtained with the recent version MAID07~\cite{MAID07} containing
many new charged-pion data from recent JLab experiments.
Figure~\ref{fig:S_waves} shows several constraints necessary for a meaningful
description of the threshold data: (I) In the Siegert limit, $\mid{\bf
{k}}\mid\rightarrow 0$ or $Q^2 \rightarrow -M_{\pi^+}^2$, the transverse
electric multipole $E_{0+}^{(-)}$ coincides with the longitudinal multipole
$L_{0+}^{(-)}$. (II) Gauge invariance requires that $\mid{\bf {k}}\mid
L_{0+}^{(-)} = k_0 S_{0+}^{(-)}$, and therefore the scalar or Coulomb multipole
$S_{0+}^{(-)}$ must be equal to the longitudinal multipole $L_{0+}^{(-)}$ at
the real photon point, $Q^2=0$. (III) Because $k_0$ vanishes for $Q^2 = 2 M_N
M_{\pi^+}+M_{\pi^+}^2 = 0.28~{\rm {GeV}}^2$, gauge invariance also implies that
the longitudinal multipole $L_{0+}^{(-)}$ decreases with $Q^2$ towards a zero
at this point. The comparison between the results based on MAID05 and MAID07
shows that $E_{0+}^{(-)}$ has not changed much, except for a change of the
slope from negative to positive, the latter being in qualitative agreement with
Eq.~(\ref{eq:E0+minus_disp}) although at a considerably smaller value. However,
the larger data base of MAID07 leads to a much smaller value for the
longitudinal multipole $L_{0+}^{(-)}$, which is now at half-way between MAID05
and the tiny value given by Eq.~(\ref{eq:L0+minus_disp}). These large
differences for the dispersive amplitudes are, however, small in view of the
huge pole terms. A comparison of Fig.~\ref{fig:S_waves} with the pole term
contributions of Eqs.~(\ref{eq:E0+minus_disp}) and (\ref{eq:L0+minus_disp})
shows that the seemingly large changes of the dispersive contributions amount
at most to a few per cent of the total amplitude. However, since the FFR sum
rule is based on the dispersive contributions, the agreement of
Eq.~(\ref{eq:multA6}) with the sum rule must be taken with a grain of salt.
%
\subsection{The pion radius and the invariant amplitude $A_5$}
%
\begin{figure*}[htb]
\begin{center}
\epsfig{figure=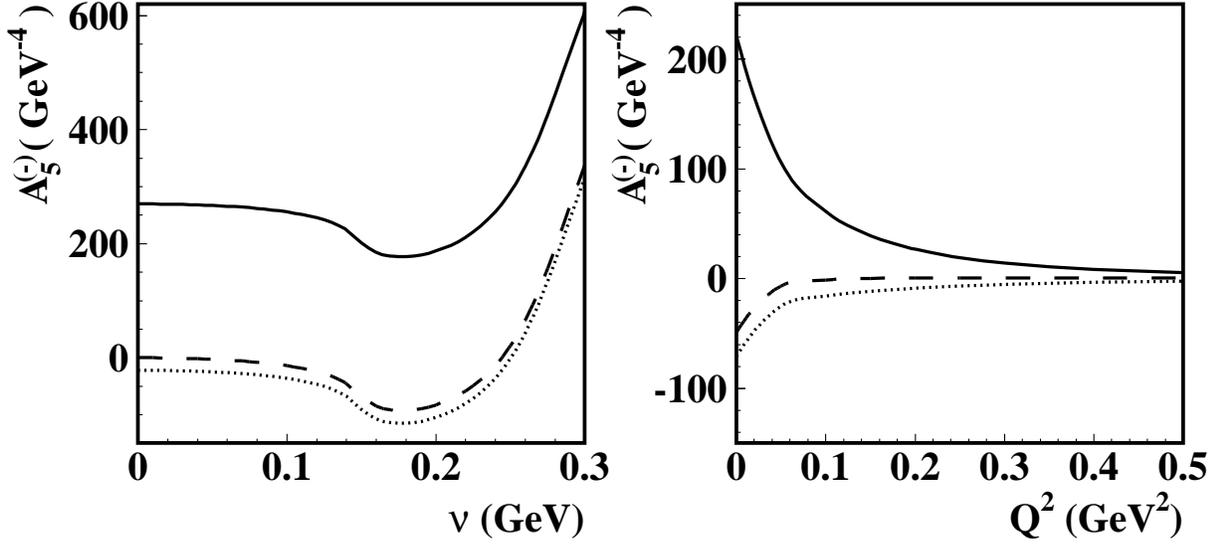,width=16.0cm,angle=0}
\end{center}
\caption{The invariant amplitude $A_{5,{\rm disp}}^{(-)}$ obtained from the
unsubtracted DR Eq.~(\ref{eq:dr1}) (solid lines), the subtracted DR of
Eq.~(\ref{eq:dr1_sub}) (dashed lines), and Eq.~(\ref{eq:A_5corr}) as proposed
by von Gehlen (dotted lines), in units of ${\rm {GeV}}^{-4}$. Left panel:
$A_{5,{\rm disp}}^{(-)}$ as function of the crossing-symmetric variable $\nu$
and at fixed values $t=t_{\rm {thr}}(0)$ and $Q^2=0$, right panel: $A_{5,{\rm
disp}}^{(-)}$ as function of the virtuality $Q^2$ at $\nu=\nu_{\rm {thr}}(Q^2)$
and $t=t_{\rm {thr}}(Q^2)$.} \label{fig:A5_sub_unsub_vG}
\end{figure*}
As pointed out by Bernard {\emph {et al.}}~\cite{Ber00}, the slope of
$L_{0+,\,{\rm disp}}^{(-)}$ given by Eq.~(\ref{eq:L0+minus_disp}),
\begin{equation}\label{eq:slope_L0loop}
\frac{d}{d\,Q^2}~L_{0+,\,{\rm disp}}^{(-)\,{\rm thr}} \approx -13.7 \cdot
\frac{10^{-3}}{{\rm{GeV}^2}\,M_{\pi^+}}\,,
\end{equation}
is comparable with the contribution of the pion radius to the slope of the pole
term, see Eq.~(\ref{eq:L0+pole_minus}) of Appendix~\ref{app:S_pole},
\begin{equation}\label{eq:slope_L0pole}
\frac{d}{d\,Q^2}~L_{0+,\,\pi~{\rm pole}}^{(-)\,{\rm thr}} \approx -22.8 \cdot
\frac{10^{-3}}{{\rm{GeV}^2}\,M_{\pi^+}}\,.
\end{equation}
Therefore, the dispersive contribution to the slope is expected to simulate a
substantial increase of the pion radius if the experimental value of  $L_{0+
}^{(-)}$ is simply compared to the pole term. In fact such an analysis yields a
modified square radius,
\begin{equation}\label{eq:eff_pi_rad}
\langle {\tilde {r}}^2 \rangle_{\pi}^V = \langle r^2 \rangle_{\pi}^V
+\frac{3}{32F_{\pi}^2}\left(\frac{16}{\pi^2}-1\right)\,.
\end{equation}
With $\langle r^2 \rangle_{\pi}^V=0.44~{\rm {fm}}^2$ as determined by
scattering high-energetic pions off atomic electrons~\cite{Ame86} and the loop
correction of 0.26~fm$^2$ (second term on the rhs of
Eq.~(\ref{eq:eff_pi_rad})), the ``effective'' value is $\langle {\tilde {r}}^2
\rangle_{\pi}^V=0.70~{\rm {fm}}^2$, corresponding to an effective pion radius
of 0.84~fm. Our dispersive analysis confirms these results, although the slope
of $L_{0+,\,{\rm disp}}^{(-)\,{\rm thr}}$ differs somewhat from
Eq.~(\ref{eq:slope_L0loop}):
\begin{eqnarray}\label{eq:slope_L0disp}
\frac{d}{d\,Q^2}~L_{0+,\,{\rm disp}}^{(-)\,{\rm thr}} =
\left\{\begin{array}{cc} -12.6 \cdot \displaystyle{\frac
{10^{-3}}{\mbox{GeV}^2\,M_{\pi^+}}}& \, \mbox{[MAID05]}\\ \;\; -9.4\cdot
\displaystyle{\frac {10^{-3}}{\mbox{GeV}^2\,M_{\pi^+}}}& \, \mbox{[MAID07]}
\end{array}\right.
\end{eqnarray}
As has been discussed before, the calculation of the amplitude $A_{5,\,{\rm
disp}}^{(-)}$ requires some care, because the unsubtracted dispersion integral
yields a contribution with the shape of the pion pole term. Since the
experimental pion form factor is already fully included in the pole term, any
additional pole structure at $\nu=0$ and $t=M_{\pi^+}^2$ ought to be removed
from $A_{5,\,{\rm disp}}^{(-)}$ according to Eq.~(\ref{eq:A_5corr}).
Figure~\ref{fig:A5_sub_unsub_vG} shows that the unsubtracted dispersion
integral (solid line) changes dramatically by removing this pole contribution
according to von Gehlen~\cite{vGe69}, see the dotted line. In fact, the
integrands for the 2 procedures show a completely different behavior as
function of the excitation energy, not only in size but also in sign. Moreover,
the unsubtracted dispersion integral does not well converge for $\nu'
\rightarrow \infty$, see also the work of Manweiler and Schmidt~\cite{Man71}
and Aznauryan~\cite{Azn03}. We have therefore corroborated the procedure of von
Gehlen by subtracting the DR at $\nu =0$ and at the given $t_{\rm {thr}}(Q^2)$,
\begin{eqnarray}
&&A_{5,{\rm disp}}^{(-)}(\nu,t,Q^2) = A_{5,{\rm disp}}^{(-)}(0,t,Q^2)\nonumber\\
&&\quad \quad +\,\frac{2\,\nu^2}{\pi}{\cal P}\int_{\nu_{thr}}^{\infty}{\rm
d}\nu' \frac{{\rm
Im}A_i^{(I)}(\nu',t,Q^2)}{\nu'(\nu'^2-\nu^2)}\,.\label{eq:dr1_sub}
\end{eqnarray}
The results are shown by the dashed lines in
Fig.~\ref{fig:A5_sub_unsub_vG}. They differ from the dotted lines by
an energy-independent shift, $A_5^{(-)}(Q^2 )\mid_{\rm {vG}} \equiv
A_{5,\,{\rm disp}}^{(-)}(\nu=0, t_{\rm {thr}}(Q^2), Q^2 )$, as
obtained from von Gehlen's procedure. In view of the excellent
convergence of the subtracted DR, Eq.~(\ref{eq:dr1_sub}), we rather
prefer this equation for further studies. In principle, the
subtraction function $A_{5,{\rm disp}}^{(-)}(0,t,Q^2)$ can be fitted
to the data by adding a polynomial in $t-M_{\pi}^2$ and $Q^2$, or by
a dispersive approach in the variable $t$. However, we have chosen
$A_5^{(-)}(Q^2 )\mid_{\rm {vG}}$ for the present discussion of the
low-energy region. The near perfect agreement with the subtracted DR
shows that the subtraction function has only a negligible dependence
on $t$ but a rapid decrease with $Q^2$, at least in the threshold
region.
\section{Summary and Discussion}
We have studied the relativistic amplitudes for pion electroproduction on the
nucleon in the framework of dispersion relations at constant $t$. This
procedure allows us to determine these amplitudes not only in the physical
region but also for sub-threshold energies. In the latter region, the
dispersive amplitudes are regular functions which can be expanded as a (real)
power series in the independent kinematic variables. A comparison of this
series with the results of ChPT yields the low-energy constants of that theory
through global properties of the excitation spectrum.\\

The present work has concentrated on the threshold region in order to compare
our findings with the predictions of several low-energy theorems based on
threshold production in the soft-pion limit. In general agreement with ChPT, we
find large corrections to these theorems due to the finite pion mass. In
particular we have studied two sum rules of Fubini~\emph {et al.}, which
connect (I) the transverse amplitude $A_1^{(N \pi^0)}$ for neutral pion
production with the Pauli form factor $F_2^N(Q^2)$ of the nucleon and (II) the
longitudinal amplitude $A_6^{(-)}$ for charged pion production with the
nucleon's axial form factor. As was shown in previous work on neutral pion
photoproduction, the former sum rule describes the anomalous magnetic moment
$\kappa_N = F_2^N(0)$ in the subthreshold region to an accuracy of a few per
cent, whereas there are sizeable unitary (rescattering) corrections at the
physical threshold. The extension to electroproduction yields large corrections
both at and below threshold, which can be visualized by pion-loop contributions
occurring at a radius of about 1.4~fm and therefore leading to a total
transition radius distinctly larger than the Pauli radius. The second sum rule
is based on the facts that (I) the pole contribution to $A_6^{(-)}$ vanishes
and (II) chiral invariance leads to the Kroll-Ruderman term as leading
threshold contribution, with the consequence that the isovector Dirac form
factor has to be replaced by the axial form factor. As a result the dispersive
amplitude for the isospin $(-)$ amplitude is proportional to $\langle
r^2\rangle_1^V-\langle r^2\rangle_A^V$ in the soft-pion limit. The dispersive
approach reproduces this sum rule at threshold, albeit at the expense of
terrific cancelations among the photoproduction multipoles stemming from
different parts of the excitation spectrum. However, the dispersive amplitude
changes rapidly if we move away from threshold, the onset of the imaginary
part, and also as function of $Q^2$.\\

We have furthermore studied the second longitudinal amplitude
$A_5^{(-)}$, which is known to converge badly. Indeed we have found
that the unsubtracted dispersion integral leads to a huge
contribution with the structure of the pion pole term, and as a
consequence to unrealistic multipoles, particularly for the
longitudinal ones. If we subtract this contribution, the threshold
amplitude changes from large positive to small negative values. In
order to corroborate this result we have also set up a subtracted
dispersion relation for $A_5^{(-)}$. The latter procedure is in
perfect agreement with the described recipe to eliminate the pion
pole structure, except for the off-set at the subtraction point. In
agreement with ChPT, our analysis yields a dispersive contribution
to the longitudinal S-wave multipole $L_{0+}^{(-)}$ with a large
negative slope in $Q^2$, which is comparable to the slope of the
pole contribution. If the experimental value of $L_{0+}^{(-)}$ is
simply compared to the pion pole term, this dispersive effect
simulates an increase of the pion
radius by about 20-25\%.\\

In conclusion, dispersion relations allow us to construct a unitary, gauge and
Lorentz invariant description of pion electroproduction on the nucleon. They
are based on the available experimental information for the absorptive parts of
the multipoles as contained, for example, in the MAID or SAID~\cite{SAID} data
analysis. In this work we have concentrated on the threshold amplitudes in
order to compare with sum rules and the predictions of ChPT. In a next step we
plan to extend our calculations to higher energies up to the $\Delta(1232)$
resonance in order to constrain the real parts of the background multipoles,
which still prevent us from a truly model-independent determination of the
small electric and Coulomb amplitudes for $\Delta(1232)$ excitation. Also along
the same lines, it is our aim to tackle the still existing discrepancies
between the electroproduction data for neutral pion production near threshold
and the respective theoretical predictions, from both ChPT and phenomenological
models like MAID. We hope that this future work will provide many new
cross-checks with relativistic effective field theories, improve our knowledge
of pion electroproduction as a means to get insight into the spatial structure
of this reaction, and lead to a better understanding of the interplay between
pion-cloud and resonance effects in the nucleon.
\section*{Acknowledgements}
This work was supported by the Deutsche Forschungsgemeinschaft (SFB 443) and
the EU Integrated Infra-structure Initiative Hadron Physics Project under
contract number RII3-CT-2004-506078.
%

\onecolumn
\section*{Appendix}
\begin{appendix}
\section{\,-\,Expansion of invariant amplitudes in terms of CGLN amplitudes}
\label{app:AtoF}
Introducing the combinations
\begin{eqnarray}
{\cal X}^{(\pm)}_{mn}  =  \frac{1}{E_i-M_N}\,{\cal F}_m \pm
\frac{E_f+M_N}{kq}\,{\cal F}_n\,, {\cal Y}^{(\pm)}_{mn}=
\frac{W-M_N}{E_i-M_N}\,{\cal F}_m \pm \frac{(W+M_N)(E_f+M_N)}{kq}\,{\cal
F}_n\,, \nonumber
\end{eqnarray}
we can cast these relations into the form
\begin{eqnarray}
\label{eq:inv_A1b} {\mathcal N}\ A_1 & = &  (W^2-M_N^2+Q^2)\,{\cal
X}^{(-)}_{12} +M_N\,\frac{k_0(t-M_\pi^2+Q^2)-2q_0Q^2}{kq}\,{\cal X}^{(+)}_{34}
+\frac{2M_NQ^2}{k_0}\, {\cal X}^{(+)}_{56} \, , \\
\label{eq:inv_A2b} {\mathcal N}\ A_2 & = & -\frac{2Q^2}{(t-M_\pi^2)}\,{\cal
X}^{(-)}_{12} +\frac{k_0(t-M_\pi^2+Q^2)-2q_0Q^2}{kq(t-M_\pi^2)}\,{\cal
Y}^{(-)}_{34}
+\frac{2Q^2}{k_0(t-M_\pi^2)}\, {\cal Y}^{(-)}_{56}\, ,  \\
\label{eq:inv_A3b} {\mathcal N}\ A_3 & = & {\cal Y}^{(+)}_{12} +
\frac{k_0(t-M_\pi^2+Q^2)-2q_0Q^2+4k^2W}{2kq}\,{\cal X}^{(+)}_{34}
+\frac{Q^2}{k_0}\,{\cal X}^{(+)}_{56}\, , \\
\label{eq:inv_A4b} {\mathcal N}\ A_4 & = & {\cal Y}^{(+)}_{12}
 +\frac{k_0(t-M_\pi^2+Q^2)-2q_0Q^2}{2kq}\,{\cal X}^{(+)}_{34}
 +\frac{Q^2}{k_0}\,{\cal X}^{(+)}_{56}\, , \\
\label{eq:inv_A5b} {\mathcal N}\ A_5 & = & \frac{t-M_\pi^2 -
4M_N\nu}{t-M_\pi^2}\,({\cal X}^{(-)}_{12} - \frac{1}{k_0}\,{\cal Y}^{(-)}_{56})
+ \frac{(t-M_\pi^2)\,(k_0+q_0-2W)+ 2M_N\nu\,(k_0-2q_0)}
{kq(t-M_\pi^2)}\,{\cal Y}^{(-)}_{34} \, , \\
\label{eq:inv_A6b} {\mathcal N}\ A_6 & = & {\cal Y}^{(+)}_{12}+2M_N\,{\cal
X}^{(-)}_{12}
 + \frac{(t-M_\pi^2+Q^2)(2W-k_0)+2(W^2-M_N^2)q_0}{2kq}\,{\cal X}^{(+)}_{34}
 - \frac{W^2-M_N^2}{k_0}\,{\cal X}^{(+)}_{56}\, ,
\end{eqnarray}
where $\mathcal{N}=W\sqrt{(E_i+M_N)(E_f+M_N)}/2\pi$.
\section{\,-\, Multipole expansion of CGLN amplitudes }
\label{app:FtoM} The multipole series of the CGLN amplitudes takes the form:
\begin{eqnarray}
\label{eq:CGLN} {\mathcal F}_1 & = & \sum_{l=0}^\infty
\,[(lM_{l+}+E_{l+})\,P_{l+1}{}'(x) +
((l+1)\,M_{l-}+E_{l-})\,P_{l-1}{}'(x)] \,, \\
{\mathcal F}_2 & = & \sum_{l=1}^\infty \, [(l+1)\,M_{l+}+lM_{l-}]\,P_l{}'(x)\,,
\\
{\mathcal F}_3 & = & \sum_{l=1}^\infty \, [(E_{l+}-M_{l+})\,P_{l+1}{}''(x) +
(E_{l-}+M_{l-})\,P_{l-1}{}''(x)]\,, \\
{\mathcal F}_4 & = & \sum_{l=2}^\infty \,
[M_{l+}-E_{l+}-M_{l-}-E_{l-}]\,P_l{}''(x)\,, \\
{\mathcal F}_5 & = & \sum_{l=0}^\infty \,
[(l+1)\,L_{l+}P_{l+1}{}'(x)-lL_{l-}P_{l-1}{}'(x)]\,, \\
{\mathcal F}_6 & = & \sum_{l=1}^\infty \, [lL_{l-}-(l+1)L_{l+}]\,P_l{}'(x)\,,
\\
{\mathcal F}_7 & = & \sum_{l=1}^\infty \, [lS_{l-}-(l+1)S_{l+}]\,P_l{}'(x)\,,
\\
{\mathcal F}_8 & = & \sum_{l=0}^\infty \,
[(l+1)S_{l+}P_{l+1}{}'(x)-lS_{l-}P_{l-1}{}'(x)]\,,
\end{eqnarray}
where $x=(t-M_{\pi}^2+Q^2+2q_0k_0)/(2qk)$ is the cosine of the scattering angle
in the physical region. The longitudinal ($L_l$) and charge ($S_l$) multipoles
are related by gauge invariance, $k_0{\mathcal F}_7=k{\mathcal F}_6$ and
$k_0{\mathcal F}_8=k{\mathcal F}_5$. In the limits of $q \rightarrow 0$
(physical threshold) and $k \rightarrow 0$ (pseudothreshold or Siegert limit),
 the multipoles have the following behavior:
 $E_{l+}, M_{l+}, L_{l+}, M_{l-} \rightarrow k^l \, q^l$ and $E_{l-}, L_{l-} \rightarrow k^{l-2} \,
 q^l$, with the exception that $L_{1-}\rightarrow k \, q $.
\section{\,-\,Multipole expansion of longitudinal amplitudes}
\label{app:A5A6}
In this appendix we give the multipole expansion of the longitudinal amplitudes
$A_5^{\rm thr}$ and $A_6^{\rm thr}$ in the notation used to describe $A_1^{\rm
thr}$ in Eqs.~(\ref{eq:thresh_A1}) and (\ref{eq:thresh_A1_alt}).
\begin{eqnarray}
A_5^{\rm thr} &=& \frac{2\pi(1+\mu)[2\mu(2+\mu)+\rho]}
{M_N^2[\mu^2(2+\mu)+\rho]\,\sqrt{(1+\mu)[(2+\mu)^2+\rho]}} \bigg\{
\frac{2(1+\mu)}{M_N[\mu^2+\rho]} E_{0+} -
\frac{4\mu(1+\mu)^2}{M_N[\mu(2+\mu)-\rho][\mu^2+\rho]}\,L_{0+}\nonumber\\
& - &
\frac{\mu(2-\mu)\sqrt{(2+\mu)^2+\rho}}{[2\mu(2+\mu)+\rho]\sqrt{\mu^2+\rho}}\,\bar{P}_2
- \frac{\mu(2+\mu)(4+8\mu+\mu^2)+(4+2\mu+\mu^2)\rho}
{[2\mu(2+\mu)+\rho]\sqrt{[(2+\mu)^2+\rho][\mu^2+\rho]
}}\,\bar{P}_3\nonumber\\
& + & \frac{2M_N(2+\mu)(2-\mu)}{2\mu(2+\mu)+\rho}\,\bar{D}
+ \frac{8(1+\mu)^2(2+\mu)}{[\mu(2+\mu)-\rho]\sqrt{[(2+\mu)^2+\rho][\mu^2+\rho]}}\,\bar{P}_5\bigg\} \nonumber\\
& = & \frac{ 2\pi (1+\mu)}
{M_N^2[\mu^2(2+\mu)+\rho]\sqrt{(1+\mu)[(2+\mu)^2+\rho]}} \,\bigg\{
-\frac{2(1+\mu)[2\mu(2+\mu)+\rho]}{M_N[\mu(2+\mu)-\rho]}\, E_{0+} \nonumber\\
&-&\frac{\mu M_N[2\mu(2+\mu)+\rho][(2+\mu)^2+\rho]}{\mu(2+\mu)-\rho}
\,\Delta_{0+}-\frac{\mu M_N(2-\mu)[(2+\mu)^2+\rho]}{2(1+\mu)}\,{\cal P}_2\nonumber\\
& - &\frac {M_N[\mu(2+\mu)(4+8\mu+\mu^2)+(4+2\mu+\mu^2)\rho]}{2(1+\mu)} \,{\cal
P}_3 +2M_N(2+\mu)(2-\mu)\,\bar{D}\nonumber\\
&+& \frac{4M_N(1+\mu)(2+\mu)[2\mu(2+\mu)+\rho]}{\mu(2+\mu)-\rho}\,{\cal P}_5
\bigg\}\,, \label{eq:thresh_A5}\\
A_6^{\rm thr}& =& \frac{4\pi}{M_N^2}\,\sqrt{\frac{1+\mu}{(2+\mu)^2+\rho}}
\bigg\{ \frac{2+\mu}{\mu^2+\rho} E_{0+} -
\frac{2\mu(1+\mu)(2+\mu)}{(\mu^2+\rho)[\mu(2+\mu)-\rho]}\,L_{0+}  + \frac{\mu
M_N}{2(1+\mu)}\,\sqrt{\frac{(2+\mu)^2+\rho}
{\mu^2+\rho}}\,\bar{P}_2\nonumber\\
&-&\frac{\mu M_N}{2(1+\mu)}\sqrt{\frac{\mu^2+\rho}{(2+\mu)^2+\rho}}\,\bar{P}_3
+ \frac{\mu M_N^2}{1+\mu}\,\bar{D}  - \frac{4 \mu M_N
(1+\mu)(2+\mu)}{[\mu(2+\mu)-\rho]\sqrt{(\mu^2+\rho)
[(2+\mu)^2+\rho]}}\,\bar{P}_5\bigg\} \nonumber\\
& = & 4\pi \, {\sqrt{\frac{1+\mu}{(2+\mu)^2+\rho}}} \,\bigg\{ -\frac
{2+\mu}{[\mu(2+\mu)-\rho]\,M_N^2}\, E_{0+} - \frac{\mu
\,(2+\mu)\,[(2+\mu)^2+\rho]}{2(1+\mu)[\mu(2+\mu)-\rho]}\, \Delta_{0+} +
\frac{\mu \,[(2+\mu)^2+\rho]}{4(1+\mu)^2} \,{\cal P}_2
\nonumber\\
& - & \frac {\mu (\mu^2+\rho)}{4(1+\mu)^2}\,{\cal P}_3  +  \frac
{\mu}{1+\mu}\,\bar{D} -\, \frac{2 \mu (2+\mu)} {\mu(2+\mu)-\rho}\,{\cal P}_5
\bigg\}\,. \label{eq:thresh_A6}
\end{eqnarray}
\section{\,-\,Pole contributions for S waves}
\label{app:S_pole} In the following we list the pole contributions and their
expansions to ${\cal O}(q^2)$ for the threshold S-wave multipoles:
\begin{eqnarray}
\label{eq:E0+pole_plusnull} E_{0+,\,{\rm pole}}^{(+,0){\rm thr}} & = &
-\frac{eg_{\pi N}}{16\pi M_N}\, \frac{[\mu\,(2+\mu)-\rho]}{(2+\mu)\,
(2+\mu+\rho)\,(1+\mu)^{3/2}}  \bigg\{ F_1^{V,S}(Q^2)+ F_2^{V/S}(Q^2)
\bigg\} \nonumber \\
& = & \frac{eg_{\pi N}}{32\pi M_N}\,( -2\mu + 3\mu^2 + \rho)(1+\kappa_{V,S}) +
\ldots \,,
\end{eqnarray}
\begin{eqnarray}
\label{eq:L0+pole_plusnull} L_{0+,\,{\rm pole}}^{(+,0){\rm thr}}  & = &
E_{0+,\,{\rm pole}}^{(+,0){\rm thr}} - \frac{eg_{\pi N}} {64\pi
M_N}\,(\mu^2+\rho)\ \frac{[\mu(2+\mu)-\rho] \, \sqrt{(2+\mu)^2+\rho}}
{(1+\mu)^{5/2}\, (2+\mu)\,(2+\mu + \rho)} \, F_2^{V,S}(Q^2) \nonumber \\
& = & \frac{eg_{\pi N}}{32\pi M_N}\,( -2\mu + 3\mu^2 + \rho)(1+\kappa_{V,S}) +
\ldots \,,
\end{eqnarray}
\begin{eqnarray}
\label{eq:E0+pole_minus}E_{0+,\,{\rm pole}}^{(-){\rm thr}}  & = & \frac{eg_{\pi
N}}{16\pi M_N}\, \frac{[(2+\mu)^2+\rho]^{3/2}}{(2+\mu)\,
(2+\mu+\rho)\,(1+\mu)^{3/2}} \nonumber \\
&& \times \bigg\{ F_1^V(Q^2) -\frac{(1+\mu)\rho}{(2+\mu)^2+\rho}\,F_2^V(Q^2)
\bigg\} \nonumber \\
& = & \frac{eg_{\pi N}}{8\pi M_N}\,\bigg\{ 1-\mu + \frac{9}{8}\mu^2 -
\frac{1}{4}\,(\kappa_V+\textstyle{\frac{1}{2}}+\textstyle{\frac{2}{3}}M^2_N
\langle r^2\rangle_1^V)\,\rho + \ldots \bigg\}\,,
\end{eqnarray}
\begin{eqnarray}
\label{eq:L0+pole_minus} L_{0+,\,{\rm pole}}^{(-){\rm thr}}  & = & E_{0+,\,{\rm
pole}}^{(-){\rm thr}} - \frac{eg_{\pi N}} {16\pi M_N}\,(\mu^2+\rho)\
\frac{\sqrt{(2+\mu)^2+\rho}}
{(1+\mu)^{3/2}\,[\mu^2(2+\mu) + \rho]} \nonumber \\
&& \times \bigg \{ F_1^V(Q^2) - \frac{[\mu^2(2+\mu) +
\rho]\,[(2+\mu)^2+\rho]}{4(2+\mu)\,
(2+\mu+\rho)\,(1+\mu)}\ F_2^V (Q^2) \nonumber \\
&& - \frac{\mu\,[\mu(2+\mu)-\rho]}{\rho(1+\mu)}\,
\bigg ( F_\pi^V(Q^2)-F_1^V(Q^2)\bigg )\bigg \} \nonumber \\
& = & E_{0+,\,{\rm pole}}^{(-){\rm thr}} + \frac{eg_{\pi N}} {8\pi
M_N}\,(\mu^2+\rho)\,\bigg\{ \frac{1}{4} \kappa_V - \frac{\sqrt{(2+\mu)^2+\rho}}
{2(1+\mu)^{3/2}\,[\mu^2(2+\mu) + \rho]}\nonumber \\
&& + \frac{\rho}{6(2\mu^2+\rho)}\,M_N^2\langle r^2\rangle_1^V + \frac{1}{6}
\left[1-\frac{\rho}{2 \mu^2+\rho }\right]\,M_N^2 \left( \langle
r^2\rangle_1^V-\langle r^2\rangle_\pi^V\right) + \ldots \bigg \}\,.
\end{eqnarray}
Strictly speaking the pion form factor can not be expanded in a power series,
because it diverges in the chiral limit. However on the phenomenological level
all the radii appearing in the above equations can be treated on the same
footing.
\section{\,-\,FFR multipoles}
\label{app:FFR} Because the FFR amplitudes are independent of $t$, the
associated current contributes only to the partial waves $0^+$ and $1^-$
corresponding to total angular momentum ${\mathcal J}=\frac{1}{2}.$ The
resulting multipole contributions are
\begin{eqnarray}
\label{eq:E0+FFR}E_{0+,\,\rm{FFR}}^{(+,0)}(W,Q^2) & = &
\sqrt{(E_i+M_N)\,(E_f+M_N)}\,\frac{W-M_N}{8\pi W}\,A_{1,\,\rm{FFR}}^{(+,0)}(Q^2)\,, \\
\label{eq:M1+FFR}\bar{M}_{1-,\,\rm{FFR}}^{(+,0)}(W,Q^2) & = &
-\sqrt{\frac{E_i-M_N}{E_f+M_N}}\,\frac{W+M_N}{8\pi W}\,A_{1,\,\rm{FFR}}^{(+,0)}(Q^2)\,, \\
\label{eq:L0+FFR}L_{0+,\,\rm{FFR}}^{(+,0)}(W,Q^2) & = &
\sqrt{(E_i+M_N)\,(E_f+M_N)}\,\frac{k_0}{8\pi W}\,A_{1,\,\rm{FFR}}^{(+,0)}(Q^2)\,, \\
\label{eq:L1+FFR}\bar{L}_{1-,\,\rm{FFR}}^{(+,0)}(W,Q^2) & = &
-\sqrt{\frac{E_i-M_N}{E_f+M_N}}\,\frac{k_0}{8\pi W}\,A_{1,\,\rm{FFR}}^{(+,0)}(Q^2)\,, \\
\label{eq:E0-FFR}E_{0+,\,\rm{FFR}}^{(-)} (W,Q^2)& = &
\sqrt{(E_i+M_N)\,(E_f+M_N)}\,\frac{Q^2}{8\pi W}\,A_{6,\,\rm{FFR}}^{(-)}(Q^2)\,, \\
\label{eq:M1-FFR}\bar{M}_{1-,\,\rm{FFR}}^{(-)}(W,Q^2) & = &
\sqrt{\frac{E_i-M_N}{E_f+M_N}}\,\frac{Q^2}{8\pi W}\,A_{6,\,\rm{FFR}}^{(-)}(Q^2)\,, \\
\label{eq:L0-FFR}L_{0+,\,\rm{FFR}}^{(-)} (W,Q^2)& = &
-\sqrt{(E_i+M_N)\,(E_f+M_N)}\,\frac{(W-M_N)\,k_0}{8\pi W}\,A_{6,\,\rm{FFR}}^{(-)}(Q^2)\,, \\
\label{eq:L1-FFR}\bar{L}_{1-,\,\rm{FFR}}^{(-)}(W,Q^2) & = &
-\sqrt{\frac{E_i-M_N}{E_f+M_N}}\,\frac{(W+M_N)\,k_0}{8\pi
W}\,A_{6,\,\rm{FFR}}^{(-)}(Q^2)\,.
\end{eqnarray}
where $A_{1,\,\rm{FFR}}^{(+,0)}$ and $A_{6,\,\rm{FFR}}^{(-)}$ are given in
Eqs.~(\ref{eq:A1FFR}) and (\ref{eq:A6FFR}), respectively. Note that the
longitudinal multipoles vanish at $k_0=0$ or $Q^2=W^2-M_N^2$, and that the
P-waves vanish at pseudo-threshold, $E_i=M_N$ or $Q^2=-(W-M_N)^2$.
%
\section{\,-\,Expansion of CGLN amplitudes in terms of invariant amplitudes }\label{app:FtoA}
The CGLN amplitudes are obtained from the invariant amplitudes by the following
equations \cite{Den61,Ber67}:
\begin{eqnarray}
\label{eq:CGLN1} \lefteqn { {\mathcal F}_1  = \frac{W-M_N}{8\pi\,W}\,
\sqrt{(E_i+M_N)(E_f+M_N)}
} \vspace{0.3cm} \nonumber \\
&& \times \bigg\{ A_1 + (W-M_N)\,A_4 - \frac{2M_N\nu_B}{W-M_N}\,(A_3-A_4) +
\frac{Q^2}{W-M_N}\,A_6\bigg\}\,,
\\
\label{eq:CGLN2} \lefteqn { {\mathcal F}_2  = \frac{W+M_N}{8\pi\,W}\,q\,
\sqrt{\frac{E_i-M_N}{E_f+M_N}}
} \vspace{0.3cm} \nonumber \\
&& \times \bigg\{ -A_1 + (W+M_N)\,A_4 - \frac{2M_N\nu_B}{W+M_N}\,(A_3-A_4)
+ \frac{Q^2}{W+M_N}\,A_6\bigg\}\,, \\
\label{eq:CGLN3} \lefteqn { {\mathcal F}_3  =
\frac{W+M_N}{8\pi\,W}\,q\,\sqrt{(E_i-M_N)(E_f+M_N)}}
\vspace{0.3cm} \nonumber \\
&& \times \bigg\{ \frac{2W^2-2M_N^2+Q^2}{2(W+M_N)}\,A_2 + A_3-A_4
- \frac{Q^2}{W+M_N}\,A_5\bigg \}\,,  \\
\label{eq:CGLN4} \lefteqn { {\mathcal F}_4  = \frac{W-M_N}{8\pi\,W}\,q^2
\,\sqrt{\frac{E_i+M_N}{E_f+M_N}}}
\vspace{0.3cm} \nonumber \\
&& \times \bigg \{ -\frac{2W^2-2M_N^2+Q^2}{2(W-M_N)} \,A_2 +A_3 - A_4
+ \frac{Q^2}{W-M_N}\,A_5\bigg \}\,, \\
\label{eq:CGLN5} \lefteqn { {\mathcal F}_5  = \frac{k_0}{8\pi W}\,
\sqrt{\frac{E_f+M_N}{E_i+M_N}} }
\vspace{0.3cm} \nonumber \\
&& \times \bigg \{ (E_i+M_N)\,A_1 +
[4M_N\nu_B(W-\textstyle{\frac{3}{4}}k_0)-k^2W+ q_0(W^2-M^2_N +
\textstyle{\frac{1}{2}}Q^2)]\,A_2+
\nonumber \\
&&[q_0(W+M_N) + 2M_N\nu_B]\,A_3 +
[(E_i+M_N)(W-M_N)-q_0(W+M_N) -2M_N\nu_B]\,A_4+  \nonumber \\
&&(2M_N\nu_Bk_0 - q_0Q^2)\,A_5 - (E_i+M_N)(W-M_N)\,A_6
\bigg\}\,,  \\
\label{eq:CGLN6} \lefteqn { {\mathcal F}_6  = \frac{k_0q}{8\pi W
\sqrt{(E_f+M_N)(E_i-M_N)}}   }
\vspace{0.3cm} \nonumber \\
&& \times \bigg \{ -(E_i-M_N)\,A_1 +
[k^2W-4M_N\nu_B)(W-\textstyle{\frac{3}{4}}k_0)-q_0(W^2-M^2_N +
\textstyle{\frac{1}{2}}Q^2)]\,A_2+
\nonumber \\
&&[q_0(W-M_N) + 2M_N\nu_B]\,A_3 +
[(E_i-M_N)(W+M_N)-q_0(W-M_N)-2M_N\nu_B]\,A_4+  \nonumber \\
&&(q_0Q^2-2M_N\nu_Bk_0)\,A_5 - (E_i-M_N)(W+M_N)\,A_6 \bigg\} \,,
\end{eqnarray}
with $\nu_B=(t-M_{\pi}^2+Q^2)/(4M_N)$.
\end{appendix}
\end{document}